\documentclass[10pt,twocolumn]{IEEEtran} %
\usepackage{amsmath,amssymb,amsfonts}
\usepackage{cite}
\usepackage{algorithm}
\usepackage{algorithmic}
\usepackage[usenames]{color}
\usepackage{subfigure}
\usepackage{url}
\usepackage{soul}
\usepackage{comment}
\usepackage{graphicx,color}
\usepackage{psfrag}
 \usepackage{epstopdf}
 \usepackage[nolist]{acronym}
\usepackage[capitalise]{cleveref}
\Crefname{equation}{Eq.\!}{Eqs.\!}
\Crefname{figure}{Fig.\!}{Figs.\!}
\Crefname{tabular}{Tab.\!}{Tabs.\!}
\Crefname{section}{Section\!}{Sections.\!}
\usepackage{pgfplots}
\usepackage{pgfplotstable}
\usepgfplotslibrary{patchplots}

\begin{document}
	
\def\th{\mathrm{th}}
\def\pt{\mathbf{p}_{_{\rm t}}}
\def\pk{\mathbf{p}_{_{ k}}}
\def\pkhat{\hat{\mathbf{p}}_{_{ k}}}
\def\pr{\mathbf{p}_{_{ \mathrm{r}, n_r}}}
\def\prf{\mathbf{p}_{_{ \mathrm{r}, 1}}}
\def\prNr{\mathbf{p}_{_{ \mathrm{r}, N_r}}}
\def\ptOne{p_{{\rm t},1}}
\def\ptTwo{p_{{\rm t},2}}
\def\prOne{p_{{\rm r},n_r,1}}
\def\prTwo{p_{{\rm r},n_r,2}}
\def\scheme{\textit{\ac{SIDCo} }} 
\def\d{d} 
\def\k{k} 
 \newcommand{\topk}{{\text{Top}_{\k}}} 
 \newcommand{\randk}{{\text{Rand}_{\k}}} 

\def\N{N}  
\def\x{\mathbf x}
\def\xin[#1]{{\x}_{\{#1\}}^{n}}
\def\gr{g} 
\def\g{\mathbf \gr}
\def\gin [#1]{{\g}_{#1}^{n}}
\def\G{G} 
\def\C{\mathbb{C}_{\k}} 
\def\Tk[#1]{{\mathbb{T}}_{\k}\left\{ #1 \right\} } 
\def\Ceta{\mathbb{C}_{\eta}} 
\def\sigmak{\sigma_{\k}}  
\def\p{p} 
\def\b{\beta} 
\def\a{\alpha} 
\def\loc{a}
\def\bhat{\hat{\beta}} 
\def\ahat{\hat{\alpha}} 
\def\lochat{\hat{a}}

\def\sign{\textrm{sign}}                                              
\def\erf{\textrm{erf}}
\def\erfc{\textrm{erfc}}

\def\Hzero{{{\mathcal{H}}_{0}}}
\def\Hone{{{\mathcal{H}}_{1}}}
\def\Hl{{{\mathcal{H}}_{l}}}

\def\Dzero{{{\mathcal{D}}_{0}}}
\def\Done{{{\mathcal{D}}_{1}}}

\def\testH1overH0{\begin{array}{l}
		{\mathcal{H}}_1 \\
		\gtrless \\
		{\mathcal{H}}_0
\end{array}}

\def\testH0overH1{\begin{array}{l}
		{\mathcal{H}}_0 \\
		\gtrless \\
		{\mathcal{H}}_1
\end{array}}

\def\testD1overD0{{ \begin{array}{l}
			{\mathcal{D}}_1 \\
			\gtrless \\
			{\mathcal{D}}_0
\end{array}}}
\newcommand{\bPhi}{\mbox{\boldmath{$\Phi$}}}
\newcommand{\Beta}{\mbox{\boldmath{$\eta$}}}
\newcommand{\bphi}{\mbox{\boldmath{$\phi$}}}
\newcommand{\bGamma}{\mbox{\boldmath{$\Gamma$}}}
\newcommand{\bgamma}{\mbox{\boldmath{$\gamma$}}}
\newcommand{\bOmega}{\mbox{\boldmath{$\Omega$}}}
\newcommand{\bomega}{\mbox{\boldmath{$\omega$}}}
\newcommand{\bbeta}{\mbox{\boldmath{$\beta$}}}
\newcommand{\bnu}{\mbox{\boldmath{$\nu$}}}
\newcommand{ \red} [ 1]{{\color{red}#1}}
\newcommand{ \blue} [ 1]{{\color{blue}#1}}
\newcommand{ \cyan} [ 1]{{\color{cyan}#1}}
\newcommand{ \magen} [ 1]{{\color{magenta}#1}}
\newcommand{ \orange} [ 1]{{\color{orange}#1}}
\newcommand{ \bfred }{\bf \color{red}}
\newcommand{ \bfblue }{\bf \color{blue}}
\newcommand{ \bfcyan} {\bf \color{cyan}}


\def\scheme{\textit{\ac{SIDCo} }} 
\def\d{d} 
\def\k{k} 

\def\N{N}  
\def\x{\mathbf x}
\def\xin[#1]{{\x}_{\{#1\}}^{n}}
\def\gr{g} 
\def\g{\mathbf \gr}
\def\gin [#1]{{\g}_{#1}^{n}}
\def\G{G} 
\def\C{\mathbb{C}_{\k}} 
\def\Tk[#1]{{\mathbb{T}}_{\k}\left\{ #1 \right\} } 
\def\Ceta{\mathbb{C}_{\eta}} 
\def\sigmak{\sigma_{\k}}  
\def\p{p} 
\def\b{\beta} 
\def\a{\alpha} 
\def\loc{a}
\def\bhat{\hat{\beta}} 
\def\ahat{\hat{\alpha}} 
\def\lochat{\hat{a}}
\def\Hbr{\mathbf{H}^{\text{BR}}\in \mathbb{C}^{N^\text{B} \times N^{\text{R}}}}
\def\Aphibr{{\textbf{A}(\boldsymbol{\phi}^\text{BR},\textbf{ r}^\text{BR})}\in \mathbb{C}^{N^\text{B} \times L^\text{BR}}}
\def\Athetabr{\textbf{A}(\boldsymbol{\theta}^\text{BR},\textbf{d}^\text{BR})\in\mathbb{C}^{N^\text{R} \times L^\text{BR}}}
\def\Hrm{\mathbf{H}^{\text{RM}}\in \mathbb{C}^{N^\text{R} \times N^{\text{M}}}}

\def\H{\mathbf{H}\in \mathbb{C}^{N^\text{B} \times N^{\text{M}}}}

\def\sign{\textrm{sign}}                                              
\def\erf{\textrm{erf}}
\def\erfc{\textrm{erfc}}

\def\Hzero{{{\mathcal{H}}_{0}}}
\def\Hone{{{\mathcal{H}}_{1}}}
\def\Hl{{{\mathcal{H}}_{l}}}

\def\Dzero{{{\mathcal{D}}_{0}}}
\def\Done{{{\mathcal{D}}_{1}}}

\def\testH1overH0{\begin{array}{l}
		{\mathcal{H}}_1 \\
		\gtrless \\
		{\mathcal{H}}_0
\end{array}}

\def\testH0overH1{\begin{array}{l}
		{\mathcal{H}}_0 \\
		\gtrless \\
		{\mathcal{H}}_1
\end{array}}

\def\testD1overD0{{ \begin{array}{l}
			{\mathcal{D}}_1 \\
			\gtrless \\
			{\mathcal{D}}_0
\end{array}}}

\def\scheme{\textit{\ac{SIDCo} }} 
\def\d{d} 
\def\k{k} 

\def\N{N}  
\def\x{\mathbf x}
\def\xin[#1]{{\x}_{\{#1\}}^{n}}
\def\gr{g} 
\def\g{\mathbf \gr}
\def\gin [#1]{{\g}_{#1}^{n}}
\def\G{G} 
\def\C{\mathbb{C}_{\k}} 
\def\Tk[#1]{{\mathbb{T}}_{\k}\left\{ #1 \right\} } 
\def\Ceta{\mathbb{C}_{\eta}} 
\def\sigmak{\sigma_{\k}}  
\def\p{p} 
\def\b{\beta} 
\def\a{\alpha} 
\def\loc{a}
\def\bhat{\hat{\beta}} 
\def\ahat{\hat{\alpha}} 
\def\lochat{\hat{a}}
\def\Hbr{\mathbf{H}^{\text{BR}}\in \mathbb{C}^{N^\text{B} \times N^{\text{R}}}}
\def\Aphibr{{\textbf{A}(\boldsymbol{\phi}^\text{BR},\textbf{ r}^\text{BR})}\in \mathbb{C}^{N^\text{B} \times L^\text{BR}}}
\def\Athetabr{\textbf{A}(\boldsymbol{\theta}^\text{BR},\textbf{d}^\text{BR})\in\mathbb{C}^{N^\text{R} \times L^\text{BR}}}
\def\Hrm{\mathbf{H}^{\text{RM}}\in \mathbb{C}^{N^\text{R} \times N^{\text{M}}}}

\def\H{\mathbf{H}\in \mathbb{C}^{N^\text{B} \times N^{\text{M}}}}

\def\sign{\textrm{sign}}                                              
\def\erf{\textrm{erf}}
\def\erfc{\textrm{erfc}}

\def\Hzero{{{\mathcal{H}}_{0}}}
\def\Hone{{{\mathcal{H}}_{1}}}
\def\Hl{{{\mathcal{H}}_{l}}}

\def\Dzero{{{\mathcal{D}}_{0}}}
\def\Done{{{\mathcal{D}}_{1}}}

\def\testH1overH0{\begin{array}{l}
		{\mathcal{H}}_1 \\
		\gtrless \\
		{\mathcal{H}}_0
\end{array}}

\def\testH0overH1{\begin{array}{l}
		{\mathcal{H}}_0 \\
		\gtrless \\
		{\mathcal{H}}_1
\end{array}}

\def\testD1overD0{{ \begin{array}{l}
			{\mathcal{D}}_1 \\
			\gtrless \\
			{\mathcal{D}}_0
\end{array}}}


\def\scheme{\textit{\ac{SIDCo} }} 
\def\d{d} 
\def\k{k} 

\def\N{N}  
\def\x{\mathbf x}
\def\xin[#1]{{\x}_{\{#1\}}^{n}}
\def\gr{g} 
\def\g{\mathbf \gr}
\def\gin [#1]{{\g}_{#1}^{n}}
\def\G{G} 
\def\C{\mathbb{C}_{\k}} 
\def\Tk[#1]{{\mathbb{T}}_{\k}\left\{ #1 \right\} } 
\def\Ceta{\mathbb{C}_{\eta}} 
\def\sigmak{\sigma_{\k}}  
\def\p{p} 
\def\b{\beta} 
\def\a{\alpha} 
\def\loc{a}
\def\bhat{\hat{\beta}} 
\def\ahat{\hat{\alpha}} 
\def\lochat{\hat{a}}
\def\Hbr{\mathbf{H}^{\text{BR}}\in \mathbb{C}^{N^\text{B} \times N^{\text{R}}}}
\def\Aphibr{{\textbf{A}(\boldsymbol{\phi}^\text{BR},\textbf{ r}^\text{BR})}\in \mathbb{C}^{N^\text{B} \times L^\text{BR}}}
\def\Athetabr{\textbf{A}(\boldsymbol{\theta}^\text{BR},\textbf{d}^\text{BR})\in\mathbb{C}^{N^\text{R} \times L^\text{BR}}}
\def\Hrm{\mathbf{H}^{\text{RM}}\in \mathbb{C}^{N^\text{R} \times N^{\text{M}}}}

\def\H{\mathbf{H}\in \mathbb{C}^{N^\text{B} \times N^{\text{M}}}}

\def\sign{\textrm{sign}}                                              
\def\erf{\textrm{erf}}
\def\erfc{\textrm{erfc}}

\def\Hzero{{{\mathcal{H}}_{0}}}
\def\Hone{{{\mathcal{H}}_{1}}}
\def\Hl{{{\mathcal{H}}_{l}}}

\def\Dzero{{{\mathcal{D}}_{0}}}
\def\Done{{{\mathcal{D}}_{1}}}

\def\testH1overH0{\begin{array}{l}
		{\mathcal{H}}_1 \\
		\gtrless \\
		{\mathcal{H}}_0
\end{array}}

\def\testH0overH1{\begin{array}{l}
		{\mathcal{H}}_0 \\
		\gtrless \\
		{\mathcal{H}}_1
\end{array}}

\def\testD1overD0{{ \begin{array}{l}
			{\mathcal{D}}_1 \\
			\gtrless \\
			{\mathcal{D}}_0
\end{array}}}

\begin{acronym}

\acro{5G-NR}{5G New Radio}
\acro{3GPP}{3rd Generation Partnership Project}
\acro{AC}{address coding}
\acro{ACF}{autocorrelation function}
\acro{ACR}{autocorrelation receiver}
\acro{ADC}{analog-to-digital converter}
\acrodef{aic}[AIC]{Analog-to-Information Converter}     
\acro{AIC}[AIC]{Akaike information criterion}
\acro{aric}[ARIC]{asymmetric restricted isometry constant}
\acro{arip}[ARIP]{asymmetric restricted isometry property}

\acro{ARQ}{automatic repeat request}
\acro{AUB}{asymptotic union bound}
\acrodef{awgn}[AWGN]{Additive White Gaussian Noise}     
\acro{AWGN}{additive white Gaussian noise}

\acro{APSK}[PSK]{asymmetric PSK} 

\acro{waric}[AWRICs]{asymmetric weak restricted isometry constants}
\acro{warip}[AWRIP]{asymmetric weak restricted isometry property}
\acro{BCH}{Bose, Chaudhuri, and Hocquenghem}        
\acro{BCHC}[BCHSC]{BCH based source coding}
\acro{BEP}{bit error probability}
\acro{BFC}{block fading channel}
\acro{BG}[BG]{Bernoulli-Gaussian}
\acro{BGG}{Bernoulli-Generalized Gaussian}
\acro{BPAM}{binary pulse amplitude modulation}
\acro{BPDN}{Basis Pursuit Denoising}
\acro{BPPM}{binary pulse position modulation}
\acro{BPSK}{binary phase shift keying}
\acro{BPZF}{bandpass zonal filter}
\acro{BSC}{binary symmetric channels}              
\acro{BU}[BU]{Bernoulli-uniform}
\acro{BER}{bit error rate}
\acro{BS}{base station}

\acro{CP}{Cyclic Prefix}
\acrodef{cdf}[CDF]{cumulative distribution function}   
\acro{CDF}{cumulative distribution function}
\acrodef{c.d.f.}[CDF]{cumulative distribution function}
\acro{CCDF}{complementary cumulative distribution function}
\acrodef{ccdf}[CCDF]{complementary CDF}               
\acrodef{c.c.d.f.}[CCDF]{complementary cumulative distribution function}
\acro{CD}{cooperative diversity}

\acro{CDMA}{Code Division Multiple Access}
\acro{ch.f.}{characteristic function}
\acro{CIR}{channel impulse response}
\acro{cosamp}[CoSaMP]{compressive sampling matching pursuit}
\acro{CR}{cognitive radio}
\acro{cs}[CS]{compressed sensing}                   
\acrodef{cscapital}[CS]{Compressed sensing} 
\acrodef{CS}[CS]{compressed sensing}
\acro{CSI}{channel state information}
\acro{CCSDS}{consultative committee for space data systems}
\acro{CC}{convolutional coding}
\acro{Covid19}[COVID-19]{Coronavirus disease}

\acro{DAA}{detect and avoid}
\acro{DAB}{digital audio broadcasting}
\acro{DCT}{discrete cosine transform}
\acro{dft}[DFT]{discrete Fourier transform}
\acro{DR}{distortion-rate}
\acro{DS}{direct sequence}
\acro{DS-SS}{direct-sequence spread-spectrum}
\acro{DTR}{differential transmitted-reference}
\acro{DVB-H}{digital video broadcasting\,--\,handheld}
\acro{DVB-T}{digital video broadcasting\,--\,terrestrial}
\acro{DL}{downlink}
\acro{DSSS}{Direct Sequence Spread Spectrum}
\acro{DFT-s-OFDM}{Discrete Fourier Transform-spread-Orthogonal Frequency Division Multiplexing}
\acro{DAS}{distributed antenna system}
\acro{DNA}{Deoxyribonucleic Acid}

\acro{EC}{European Commission}
\acro{EED}[EED]{exact eigenvalues distribution}
\acro{EIRP}{Equivalent Isotropically Radiated Power}
\acro{ELP}{equivalent low-pass}
\acro{eMBB}{Enhanced Mobile Broadband}
\acro{EMF}{electric and magnetic fields}
\acro{EU}{European union}

\acro{FC}[FC]{fusion center}
\acro{FCC}{Federal Communications Commission}
\acro{FEC}{forward error correction}
\acro{FFT}{fast Fourier transform}
\acro{FH}{frequency-hopping}
\acro{FH-SS}{frequency-hopping spread-spectrum}
\acrodef{FS}{Frame synchronization}
\acro{FSsmall}[FS]{frame synchronization}  
\acro{FDMA}{Frequency Division Multiple Access}

\acro{GA}{Gaussian approximation}
\acro{GF}{Galois field }
\acro{GG}{Generalized-Gaussian}
\acro{GIC}[GIC]{generalized information criterion}
\acro{GLRT}{generalized likelihood ratio test}
\acro{GPS}{Global Positioning System}
\acro{GMSK}{Gaussian minimum shift keying}
\acro{GSMA}{Global System for Mobile communications Association}

\acro{HAP}{high altitude platform}

\acro{IDR}{information distortion-rate}
\acro{IFFT}{inverse fast Fourier transform}
\acro{iht}[IHT]{iterative hard thresholding}
\acro{i.i.d.}{independent, identically distributed}
\acro{IoT}{Internet of Things}                      
\acro{IR}{impulse radio}
\acro{lric}[LRIC]{lower restricted isometry constant}
\acro{lrict}[LRICt]{lower restricted isometry constant threshold}
\acro{ISI}{intersymbol interference}
\acro{ITU}{International Telecommunication Union}
\acro{ICNIRP}{International Commission on Non-Ionizing Radiation Protection}
\acro{IEEE}{Institute of Electrical and Electronics Engineers}
\acro{ICES}{IEEE international committee on electromagnetic safety}
\acro{IEC}{International Electrotechnical Commission}
\acro{IARC}{International Agency on Research on Cancer}
\acro{IS-95}{Interim Standard 95}

\acro{LEO}{low earth orbit}
\acro{LF}{likelihood function}
\acro{LLF}{log-likelihood function}
\acro{LLR}{log-likelihood ratio}
\acro{LLRT}{log-likelihood ratio test}
\acro{LOS}{Line-of-Sight}
\acro{LRT}{likelihood ratio test}
\acro{wlric}[LWRIC]{lower weak restricted isometry constant}
\acro{wlrict}[LWRICt]{LWRIC threshold}
\acro{LPWAN}{low power wide area network}
\acro{LoRaWAN}{Low power long Range Wide Area Network}
\acro{NLOS}{non-line-of-sight}

\acro{MB}{multiband}
\acro{MC}{multicarrier}
\acro{MDS}{mixed distributed source}
\acro{MF}{matched filter}
\acro{m.g.f.}{moment generating function}
\acro{MI}{mutual information}
\acro{MIMO}{multiple-input multiple-output}
\acro{MISO}{multiple-input single-output}
\acrodef{maxs}[MJSO]{maximum joint support cardinality}                       
\acro{ML}[ML]{maximum likelihood}
\acro{MLE}[ML]{maximum likelihood}
\acro{MMSE}{minimum mean-square error}
\acro{MMV}{multiple measurement vectors}
\acrodef{MOS}{model order selection}
\acro{M-PSK}[${M}$-PSK]{$M$-ary phase shift keying}                       
\acro{M-APSK}[${M}$-PSK]{$M$-ary asymmetric PSK} 

\acro{M-QAM}[$M$-QAM]{$M$-ary quadrature amplitude modulation}
\acro{MRC}{maximal ratio combiner}                  
\acro{maxs}[MSO]{maximum sparsity order}                                      
\acro{M2M}{machine to machine}                                                
\acro{MUI}{multi-user interference}
\acro{mMTC}{massive Machine Type Communications}      
\acro{mm-Wave}{millimeter-wave}
\acro{MP}{mobile phone}
\acro{MPE}{maximum permissible exposure}
\acro{MAC}{media access control}
\acro{NB}{narrowband}
\acro{NBI}{narrowband interference}
\acro{NLA}{nonlinear sparse approximation}
\acro{NLOS}{Non-Line of Sight}
\acro{NTIA}{National Telecommunications and Information Administration}
\acro{NTP}{National Toxicology Program}
\acro{NHS}{National Health Service}

\acro{OC}{optimum combining}                             
\acro{OC}{optimum combining}
\acro{ODE}{operational distortion-energy}
\acro{ODR}{operational distortion-rate}
\acro{OFDM}{orthogonal frequency-division multiplexing}
\acro{omp}[OMP]{orthogonal matching pursuit}
\acro{OSMP}[OSMP]{orthogonal subspace matching pursuit}
\acro{OQAM}{offset quadrature amplitude modulation}
\acro{OQPSK}{offset QPSK}
\acro{OFDMA}{Orthogonal Frequency-division Multiple Access}
\acro{OPEX}{Operating Expenditures}
\acro{OQPSK/PM}{OQPSK with phase modulation}

\acro{PAM}{pulse amplitude modulation}
\acro{PAR}{peak-to-average ratio}
\acrodef{pdf}[PDF]{probability density function}                      
\acro{PDF}{probability density function}
\acrodef{p.d.f.}[PDF]{probability distribution function}
\acro{PDP}{power dispersion profile}
\acro{PMF}{probability mass function}                             
\acrodef{p.m.f.}[PMF]{probability mass function}
\acro{PN}{pseudo-noise}
\acro{PPM}{pulse position modulation}
\acro{PRake}{Partial Rake}
\acro{PSD}{power spectral density}
\acro{PSEP}{pairwise synchronization error probability}
\acro{PSK}{phase shift keying}
\acro{PD}{power density}
\acro{8-PSK}[$8$-PSK]{$8$-phase shift keying}

\acro{FSK}{frequency shift keying}

\acro{QAM}{Quadrature Amplitude Modulation}
\acro{QPSK}{quadrature phase shift keying}
\acro{OQPSK/PM}{OQPSK with phase modulator }

\acro{RD}[RD]{raw data}
\acro{RDL}{"random data limit"}
\acro{ric}[RIC]{restricted isometry constant}
\acro{rict}[RICt]{restricted isometry constant threshold}
\acro{rip}[RIP]{restricted isometry property}
\acro{ROC}{receiver operating characteristic}
\acro{rq}[RQ]{Raleigh quotient}
\acro{RS}[RS]{Reed-Solomon}
\acro{RSC}[RSSC]{RS based source coding}
\acro{r.v.}{random variable}                               
\acro{R.V.}{random vector}
\acro{RMS}{root mean square}
\acro{RFR}{radiofrequency radiation}
\acro{RIS}{Reconfigurable Intelligent Surface}
\acro{RNA}{RiboNucleic Acid}

\acro{SA}[SA-Music]{subspace-augmented MUSIC with OSMP}
\acro{SCBSES}[SCBSES]{Source Compression Based Syndrome Encoding Scheme}
\acro{SCM}{sample covariance matrix}
\acro{SEP}{symbol error probability}
\acro{SG}[SG]{sparse-land Gaussian model}
\acro{SIMO}{single-input multiple-output}
\acro{SINR}{signal-to-interference plus noise ratio}
\acro{SIR}{signal-to-interference ratio}
\acro{SISO}{single-input single-output}
\acro{SMV}{single measurement vector}
\acro{SNR}[\textrm{SNR}]{signal-to-noise ratio} 
\acro{sp}[SP]{subspace pursuit}
\acro{SS}{spread spectrum}
\acro{SW}{sync word}
\acro{SAR}{specific absorption rate}
\acro{SSB}{synchronization signal block}

\acro{TH}{time-hopping}
\acro{ToA}{time-of-arrival}
\acro{TR}{transmitted-reference}
\acro{TW}{Tracy-Widom}
\acro{TWDT}{TW Distribution Tail}
\acro{TCM}{trellis coded modulation}
\acro{TDD}{time-division duplexing}
\acro{TDMA}{Time Division Multiple Access}

\acro{UAV}{unmanned aerial vehicle}
\acro{uric}[URIC]{upper restricted isometry constant}
\acro{urict}[URICt]{upper restricted isometry constant threshold}
\acro{UWB}{ultrawide band}
\acro{UWBcap}[UWB]{Ultrawide band}   
\acro{URLLC}{Ultra Reliable Low Latency Communications}
         
\acro{wuric}[UWRIC]{upper weak restricted isometry constant}
\acro{wurict}[UWRICt]{UWRIC threshold}                
\acro{UE}{user equipment}
\acro{UL}{uplink}

\acro{WiM}[WiM]{weigh-in-motion}
\acro{WLAN}{wireless local area network}
\acro{wm}[WM]{Wishart matrix}                               
\acroplural{wm}[WM]{Wishart matrices}
\acro{WMAN}{wireless metropolitan area network}
\acro{WPAN}{wireless personal area network}
\acro{wric}[WRIC]{weak restricted isometry constant}
\acro{wrict}[WRICt]{weak restricted isometry constant thresholds}
\acro{wrip}[WRIP]{weak restricted isometry property}
\acro{WSN}{wireless sensor network}                        
\acro{WSS}{wide-sense stationary}
\acro{WHO}{World Health Organization}
\acro{Wi-Fi}{wireless fidelity}

\acro{sss}[SpaSoSEnc]{sparse source syndrome encoding}

\acro{VLC}{visible light communication}
\acro{VPN}{virtual private network} 
\acro{RF}{radio frequency}
\acro{FSO}{free space optics}
\acro{IoST}{Internet of space things}

\acro{GSM}{Global System for Mobile Communications}
\acro{2G}{second-generation cellular network}
\acro{3G}{third-generation cellular network}
\acro{4G}{fourth-generation cellular network}
\acro{5G}{5th-generation cellular network}	
\acro{gNB}{next generation node B base station}
\acro{NR}{New Radio}
\acro{UMTS}{Universal Mobile Telecommunications Service}
\acro{LTE}{Long Term Evolution}
\acro{QoS}{Quality of Service}
\acro{Tx}{transmitter}
\acro{Rx}{receiver}
\acro{AoA}{angle of arrival}
\acro{AoD}{angle of departure}
\acro{TDoA}{time difference of arrival}
\acro{ToA}{time of arrival}
\acro{LS}{least square}
\acro{CRLB}{Cramer-Rao lower bound}
\acro{EKF}{extended Kalman filtering}
\acro{MSE}{mean square error}
\acro{P-CRLB}{posterior CRLB}
\acro{RSS}{received signal strength}
\acro{MAP}{maximum a posteriori}
\acro{LoS}{line of sight}
\acro{PEB}{Position Error Bound}
\end{acronym}

\title{ Single Antenna Tracking and Localization of RIS-enabled Vehicular Users}
\author{Somayeh~Aghashahi,~\IEEEmembership{Student Member,~IEEE,}
        Zolfa~Zeinalpour-Yazdi,~\IEEEmembership{Member,~IEEE,}
        Aliakbar~Tadaion,~\IEEEmembership{Senior Member,~IEEE,}
      Mahdi~Boloursaz Mashhadi,~\IEEEmembership{Senior Member,~IEEE,}
        and Ahmed~Elzanaty,~\IEEEmembership{Senior Member,~IEEE}

\thanks{Copyright (c) 2024 IEEE. Personal use of this material is permitted. However, permission to use this material for any other purposes must be obtained from the IEEE by sending a request to pubs-permissions@ieee.org.}
\thanks{This work was supported in part by Iran National Science Foundation (INSF) under project No. 4013201 and in part by the Towards IoT Everywhere Project (KAUST-Surrey-Maynooth) under the Award ORFS-2022-CRG11-5058.2.}
     \thanks{S. Aghashahi, Z. Zeinalpour-Yazdi and A. Tadaion are with  the Department of Electrical Engineering, Yazd University, Yazd, Iran (e-mail: aghashahi@stu.yazd.ac.ir and \{zeinalpour, tadaion\}@yazd.ac.ir).}
     \thanks{M. B. Mashhadi  and A. Elzanaty  are with 5GIC \& 6GIC, Institute for Communication Systems (ICS), University of Surrey, Guildford, GU2 7XH, United Kingdom (e-mail: \{m.boloursazmashhadi, a.elzanaty\}@surrey.ac.uk). }
     }
\maketitle
\begin{abstract}
\acp{RIS}  are envisioned to be employed in next generation wireless networks to enhance the communication and radio localization services. 
In this paper, we propose  novel localization and tracking algorithms exploiting reflections through \acp{RIS} at multiple receivers. We utilize a  single antenna \ac{Tx} and multiple single antenna \acp{Rx} to estimate the position and the velocity of users (e.g. vehicles) equipped with \acp{RIS}. 
Then, we design the \ac{RIS} phase shifts to separate the signals from different users.
The proposed algorithms exploit the geometry information  of the signal at the \acp{RIS} to localize and track the users. 
  We also conduct a comprehensive analysis of the \ac{CRLB}  of the localization  system.
Compared to the \ac{ToA}-based localization approach, the proposed method reduces the localization error   by a factor up to three. Also, the simulation results show the accuracy of the proposed tracking approach.
\end{abstract}
	\acresetall 

\section{Introduction}
\IEEEPARstart{W}{ith}  the progress of technology over the last decades, localization and tracking have become one of the necessities in human life. Applications of localization are in  navigation of vehicles and
robots, emergency services, location-based services in wireless networks, security, etc.
Due to the inherent limitations of \ac{GPS}  in indoor environments and outdoor environments with tall buildings, the attention of the researchers has been turned to the radio localization, motivating a huge interest in the integration of localization, sensing and communication (ISAC) in future wireless systems \cite{liu2022integrated,wei2023integrated,9330512,10172231,10174742}.
{ The main challenge of radio localization systems is to simultaneously achieve wide coverage  and high accuracy, while using the minimum number of \acp{Tx}/\acp{Rx} \cite{1Gto5G,elzanaty2023towards}}. A promising technology to help overcome this challenge is \ac{RIS}.

An \ac{RIS} is a planner array composed of passive reflecting elements with dynamically configurable phase shifts.
 \acp{RIS} have been used not only to enhance wireless communication performance \cite{wu2019towards,liu2021reconfigurable,10345757,10066522}, but also to boost radio localization capabilities \cite{wymeersch2020radio,keykhosravi_VTM}. In many of the localization researches the \acp{RIS} are considered as anchor nodes with known location. 
The deployment of \acp{RIS}  to improve the accuracy is one of the initial problems explored in this context. 
 In this regard, enhancing the localization accuracy in a system equipped by a single antenna \ac{Tx}/\ac{Rx} and multi-antenna \ac{Rx}/\ac{Tx} has been examined in \cite{UWB,10123105,10557757,wu2024employing}. Also, in \cite{orientation,rinchi2022compressive,beamtrain},  
   this problem is investigated in a system equipped with multi-antenna \ac{Tx} and \ac{Rx}. Moreover, the authors in \cite{SISO3D,LOS_NOLS, AOI,zhang2020towards} investigated the problem of localizing a single antenna \ac{UE} employing a single antenna \ac{Tx}. Another challenge is employing \acp{RIS} to localize the  \acp{UE} without any access point \cite{zero_access,zero_access_SLAM}. Also, the problem of localizing unknown location \acp{Tx} and \acp{Rx} has been investigated in \cite{chen2023multi,ammous2023zero}.  The enhancement of the tracking accuracy is another role for the \acp{RIS}, which is studied in \cite{EKF_tacking,platoon}.
   
However, all of these works assume that the location of the \ac{RIS} is known and use it as anchor. But, what if the \acp{RIS} are deployed on vehicles and we are uncertain about their precise location? In such situations, we need to estimate the location of the \ac{RIS} itself. 
There are a limited number of  studies that address the \ac{RIS} localization problem  \cite{semipassive,bi-static,calibration1,calibration2,calibration3}. 
In \cite{semipassive}, localization of \ac{RIS}-equipped \ac{UE}s using a \ac{Tx} and multiple \acp{Rx} is considered, where, for each \ac{RIS} the \ac{TDoA} of the path including the \ac{RIS} and each \ac{Rx} and the direct path between the \ac{Tx} and \ac{Rx} is measured and the location of the \ac{RIS} is estimated by minimizing the \ac{ML} for the \ac{TDoA}s. Thus, at least three \acp{Rx} are required to localize the \acp{RIS} in this work. 
In \cite{bi-static}, the problem of localizing an \ac{RIS}  in the near field is investigated. In this article,  
a \ac{ML} estimation problem is defined to localize the \ac{RIS}
using one \ac{Tx} and one \ac{Rx}.
 In \cite{calibration1,calibration2,calibration3}, a \ac{MIMO} \ac{Rx} is employed to estimate the location of the RIS and the \ac{Tx} (user). To do this, in \cite{calibration1} a prior information about the locations is utilized, while in \cite{calibration2,calibration3} active and hybrid \acp{RIS} are employed, respectively.

However, these works mostly consider static scenarios which is not suitable for mobile RIS such as vehicles equipped with RIS in smart transportation systems where the  mobility features such as Doppler effect can not be ignored.To the best of our knowledge, this paper is the first study of the \ac{RIS} tracking problem.

\begin{figure}[t!]
\includegraphics[width=\columnwidth]{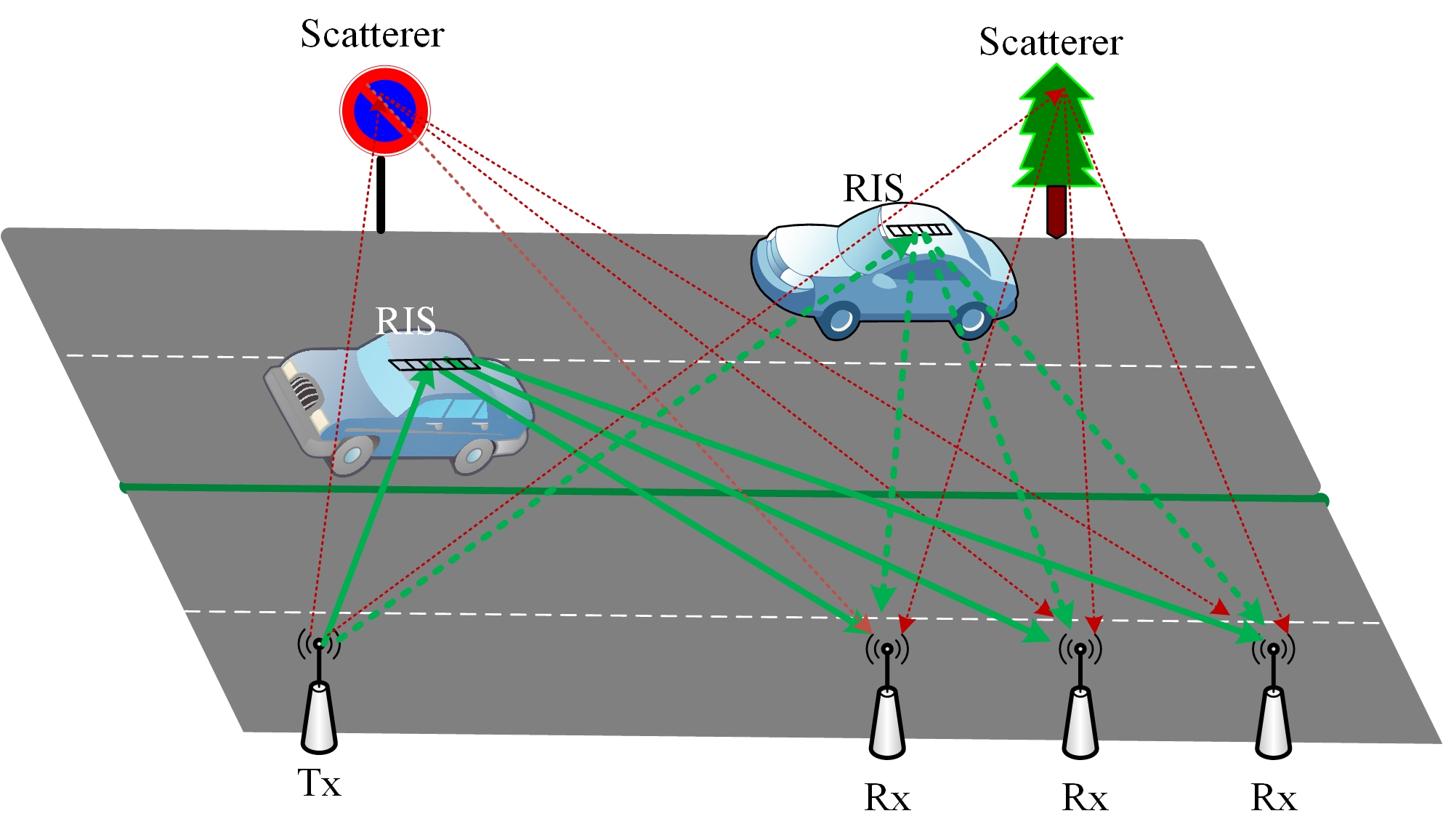}
\caption{A capture of the considered scenario, as $N_{\mathrm{r}}=3$, $K=2$ and $L=2$.}
\label{general_scenario}
\end{figure}
\begin{table*}[!t]
\caption{The Key Differences Between This Work and the Literature }
\label{tab:Key_diff}
\vspace{-2mm}
\centering
{
\begin{tabular}{|c|c|c|c|c|c|c|c|c|}
\hline
& Tracking &	\shortstack{Single Antenna \\ Tx and Rx} &  \shortstack{Doppler \\ Effect} & \shortstack{Simultaneous Tracking \\/Localization  of Multiple Targets} & \shortstack{CRLB \\ Analysis} &  Scatterers	 & \shortstack{Unknown \\ Location RIS} &\shortstack{ Unknown \\ Location Tx/Rx}										\\ \hline\hline
\cite{UWB} & N & N  & N & N & Y &	N	& N & Y	\\ \hline
\cite{finger_multi_antanna} & N  & N  & N  & Y & N &		Y & N & Y	\\ \hline
\cite{beamtrain} & N & N & N & N & Y  &	Y	& N & Y	\\ \hline
\cite{platoon} & Y & N &  N &Y & Y &	 N	& N & Y	\\ \hline
\cite{bi-static}& N & Y & N &N & Y &	N	& Y & N		\\ \hline
\cite{semipassive} & N & Y & N & Y &  Y&	  N    & Y & N	                                                       \\ \hline
 \cite{calibration1} & N &N &N &Y &Y &N	& Y & Y \\ \hline
 \cite{calibration2}  &N &N &N &N &Y &	N & Y & Y  \\ \hline
\cite{calibration3}    & N &N &N &N &Y &N	 & Y & Y	\\ \hline
This work & Y & Y & Y & Y & Y & Y	 & Y & N		\\ \hline
\end{tabular}}
\end{table*}
In this paper, we study the problem of  tracking the trajectory and speed of multiple \ac{UE}s (e.g., vehicles) each equipped with an \ac{RIS}, in the presence of the scatters,  employing a single antenna \ac{Tx} and multiple single antenna \acp{Rx}. In this regard, first we propose a novel localization algorithm to estimate the initial location of the \acp{RIS}. Then, we propose an efficient algorithm for simultaneous tracking of multiple \acp{RIS}. Table \ref{tab:Key_diff} summarizes the key differences between this work and the literature.
 The contributions of this paper are summarized as follows.
\begin{itemize}
\item We propose an  algorithm  to simultaneously track the trajectory and speed of multiple \acp{RIS} based on \ac{EKF}. The proposed approach leverages the \ac{ToA} of the paths and  information about the geometry of the reflected signals at the \ac{RIS}. 
\item 
By estimating the \acp{ToA} and sum cosines of the \ac{AoA} and \ac{AoD} at each \ac{RIS}  as the measurement parameters, we also propose an \ac{RIS} localization algorithm  to initialize the tracking. To do so, we design the phase shifts to extract the signals reflected by various \acp{RIS} and discriminate them from those reflected by scatterers at each \ac{Rx}.
Thanks to the exploited geometric information, the proposed localization approach works even with a single \ac{Rx}. 
\item
We derive the \ac{CRLB}  
and demonstrate the behaviour of \ac{PEB} in terms of the distance between the \ac{RIS} and the line between the \ac{Tx} and \ac{Rx}.
\end{itemize}

 The rest of the paper is organized as follows.
 In Section \ref{sec:system_model}, we present the system model. In Section \ref{sec:localization}, the phase shifts design approach, measurement parameters and the proposed localization approach are introduced. Section \ref{sec:tracking} provides the proposed tracking approach.  
The \ac{CRLB} of the localization system is derived in Section \ref{sec:CRLB}.
  Section \ref{sec:simulation} presents the simulation results, and Section \ref{sec:conclusion} concludes the paper.
 
\textbf{{Notations.}}
The vectors  and the matrices  are denoted with boldface letters and
boldface capital letters, respectively. For matrix $\bf A$, $\mathrm{tr}\{\mathbf{A}\}$ denotes the trace of the matrix and $\mathbf{A}^{T}$, $\mathbf{A}^{H}$, $\mathbf{A}^*$,  and  $[\bf A]_{i,j}$ are the transpose, Hermitian, conjugate and element $(i,j)$ of the matrix, respectively. Also, for vector $\mathbf{u}$, $[\mathbf{u}]_i$ shows the $i^{\th}$ element of this vector, and for vectors $\mathbf{u}$ and $\mathbf{v}$,
$\mathbf{u}\odot \mathbf{v}$ represents the element-wise multiplication of the vectors. 
Moreover, operators $| . |$ and $\parallel. \parallel$ return the  absolute value of {scalars} and the norm of   vectors, respectively.

\section{System Model}
\label{sec:system_model}
\begin{figure}[t!]
\psfrag{Pt}{$\pt$}
\psfrag{Pr}{$\pr$}
\psfrag{Pk}{$\pk$}
\psfrag{phik}{$\phi_k$}
\psfrag{pithetak}{$\pi - \theta_{k, n_{\mathrm{r}}}$}
\includegraphics[width=\columnwidth]{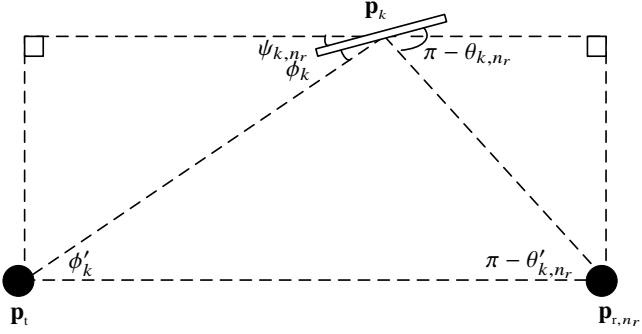}
\caption{Representation of locations and \ac{AoA} and \ac{AoD}   in a scenario including the \ac{Tx}, the $n^{\th}$ \ac{Rx}, and the $k^{\th}$ \ac{RIS}.}
\label{parallel_scenario}
\end{figure}

We consider a localization system comprising a single antenna \ac{Tx} and $N_{\mathrm{r}}$ single antenna \acp{Rx} deployed in an environment with $L$ scatterers (as in Fig. \ref{general_scenario}). We consider $K$  users (e.g., vehicles) with unknown locations, each {equipped with a linear \ac{RIS}\cite{linear_RIS_1,linear_RIS_2,linear_RIS_3,linear_RIS_4}}, whose locations trajectory and speed are to be estimated. 
 We denote the location of the \ac{Tx}, the $n_{\mathrm{r}}^{\th}$ \ac{Rx} and  $k^{\th}$ \ac{RIS} with $\pt$, $\pr$, and $\pk$, respectively, as shown in Fig.~\ref{parallel_scenario}.
 { Moreover, $\phi_k$, $\theta_{k,n_{\mathrm{r}}}$ and $\psi_{k,n_{\mathrm{r}}}$  represent the  \ac{AoA} of the path between the \ac{Tx} and the $k^{\th}$ \ac{RIS} and the \ac{AoD} of the path between the $k^{\th}$ \ac{RIS} and the $n_{\mathrm{r}}^{\th}$ \ac{Rx},  and the orientation of the $k^{\th}$ \ac{RIS} with respect to the line between the \ac{Tx} and the $n_{\mathrm{r}}^{\th}$ \ac{Rx},   respectively.} Furthermore, the steering vectors of the \ac{AoA} and \ac{AoD}s of the $k^{\th}$ \ac{RIS} are denoted as $\mathbf{a}(\phi_k)$ and $\mathbf{a}(\theta_{k,n_{\mathrm{r}}})$ for $n_{\mathrm{r}} \in \mathcal{N}_{\mathrm{r}}\triangleq\{1,\dots,n_{\mathrm{r}}\}$, where 
 { 
\begin{equation}
\mathbf{a}(\psi)=[1, e^{j \frac{2\pi}{\lambda}d \cos \psi},  \dots, e^{j \frac{2\pi}{\lambda}(M-1) d \cos \psi}]^{T},
\end{equation}}
in which $M$ is the number of the \ac{RIS} elements.

The \ac{Tx} sends an \ac{OFDM} signal with $N$ sub-carriers  for $\bar{T}$ times. At the $n_{\mathrm{r}}^{\th}$ \ac{Rx}, after removing  the \ac{CP} and applying \ac{FFT}, the  received signal  can be written as
\begin{align}
 \notag \hspace{-1mm}\mathbf{y}_{n_{\mathrm{r}},\bar{t}}\hspace{-1mm}=&\sqrt{E_s}\sum\limits_{k=1}^{K} g_{k,n_{\mathrm{r}}} \mathbf{d}(\tau_{k,n_{\mathrm{r}}}) \mathbf{a}(\theta_{k,n_{\mathrm{r}}})^T\bPhi_{k}(\bar{t})\mathbf{a}(\phi_k)\\ & \times e^{j2\pi f_{_{\mathrm{d},k,n_{\mathrm{r}}}} T_{\mathrm{d}} (\bar{t}-1)  }
 +\sqrt{E_s}\sum\limits_{\ell=0}^{L} {g}^{\prime}_{\ell,n_{\mathrm{r}}} \mathbf{d}({\tau}^{\prime}_{\ell, n_{\mathrm{r}}}) +\mathbf{w}_{n_{\mathrm{r}},\bar{t}}, \notag \\ &  \forall \bar{t}\in \{1,2,\cdots,\bar{T}\}, \label{eq:y_t_bar}
\end{align}
where $E_s$ is the energy of the transmit signal, $$\mathbf{d}(\tau)=[1, e^{j 2 \pi \Delta_f \tau}, \dots, e^{j 2 \pi (N-1) \Delta_f \tau}]^{T},$$ $\bPhi_{k}(\bar{t})\in \mathbb{C}^{M\times M}$ is the phase profile matrix  of the $k^{\th}$ \ac{RIS} at the $\bar{t}^{\th}$ transmission, and $\mathbf{w}_{n_{\mathrm{r}},\bar{t}}$ is the \ac{AWGN} vector at the $n_{\mathrm{r}}^{\th}$  \ac{Rx} with power spectral density $N_0$.
Also,
 $g_{k,n_{\mathrm{r}}}$ and $\tau_{k,n_{\mathrm{r}}}$  are   the channel amplitude and  the delay of the path between the \ac{Tx} and the $n_{\mathrm{r}}^{\th}$ \ac{Rx} including the $k^{\th}$ \ac{RIS}, respectively, and ${g}^{\prime}_{\ell,n_{\mathrm{r}}}
$ and ${\tau}^{\prime}_{\ell,n_{\mathrm{r}}} $ are  the channel amplitude and the delay of the path between the \rm{Tx} and the $n_{\mathrm{r}}^{\th}$ \ac{Rx} including the $\ell^{th \rm}$ scatterer, respectively  ($\ell=0$ is corresponding to the direct paths between the \ac{Tx} and \acp{Rx}).
{Additionally,  $f_{_{\mathrm{d},k,n_{\mathrm{r}}}}=v_k(\cos \phi_{k,n_{\mathrm{r}}}+ \cos \theta_{k,n_{\mathrm{r}}})f_c/c$ is the Doppler frequency of the signal reflected by the $k^{\th}$ \ac{RIS} and received by the $n_{\mathrm{r}}^{\th}$ \ac{Rx}  in which $v_k$ is the velocity of the $k^{\th}$ \ac{RIS}, $f_c$ is the carrier frequency, $c$ is the speed of light, and $T_d$ is the symbol duration of the \ac{OFDM} signal plus duration of the \ac{CP}.}

\section{Proposed Localization Approach}
\label{sec:localization}
In order to investigate the tracking problem, first we need to consider measurement parameters and also design a localization approach to estimate the initial location of the \acp{RIS}. 
  In this regard,  in this section, we propose a phase shift design approach and estimate the \ac{ToA} and the Doppler frequency of the paths between the \ac{Tx} and \acp{Rx} including the $k^{\th}$ \ac{RIS} as well as parameter  $\alpha_{k,n_{\mathrm{r}}}\triangleq\cos \phi_k+\cos \theta_{k,n_{\mathrm{r}}}$, for $k\in \mathcal{K}\triangleq\{1,\dots, K\}$ and $n_{\mathrm{r}} \in \mathcal{N}_{\mathrm{r}}$. Then, using these parameters we define a \ac{LS} problem to localize the $k^{\th}$ \ac{RIS}. We also will employ the \ac{ToA}s and $\alpha_{k,n_{\mathrm{r}}}$s as the measurement parameters to propose the tracking algorithm in the next section.

\subsection{Phase shift and observation vector design}
We have three purposes  in the design of RIS phase shifts: (i) Be able to eliminate the effect of scatterers. (ii) Be able to separate the signals of different RISs. (iii) Be able to estimate the value of $\alpha_{k,n_{\mathrm{r}}}$.
In this part, we design the phase shifts of the \acp{RIS} in order to separate the signals reflected
by various \acp{RIS} at the \acp{Rx} and discriminate them from the scatterers. This enables simultaneous localization and tracking of multiple \acp{RIS} from the received signals.
 To do this, we divide $\bar{T}$ into $N_T$ intervals each including  $T$ slots (i.e., $\bar{T}=N_T T$) and for $n_T\in \mathcal{N}_T\triangleq\{1,\dots, N_T\}$  consider $\mathbf{Y}_{n_{\mathrm{r}}, n_T}\in \mathbb{C}^{N \times T}$ as the matrix form of the signals received by the $n_{\mathrm{r}}^{\th}$ \ac{Rx} at the slots of this interval. Thus, 
\begin{align}
\notag \mathbf{Y}_{n_{\mathrm{r}}, n_T}=&\sqrt{E_s}\sum\limits_{k=1}^{K} g_{k,n_{\mathrm{r}}} e^{j 2\pi f_{_{\mathrm{d},k,n_{\mathrm{r}}}} (n_T-1) T_{\mathrm{d}} T} \\ &  \times \mathbf{d}(\tau_{k,n_{\mathrm{r}}}) \notag [\mathbf{u}_{n_T}(\theta_{k, n_{\mathrm{r}}},\phi_k)  \odot \mathbf{d}_\mathrm{d}(f_{_{\mathrm{d},k,n_{\mathrm{r}}}})]^T   \\ & +\sqrt{E_s}\sum\limits_{\ell=0}^{L} g_{\ell,n_{\mathrm{r}}}^{\prime} \mathbf{d}(\tau_{\ell,n_{\mathrm{r}}}^{\prime}) 1_{T}^{T}+\mathbf{W}_{n_{\mathrm{r}},n_T}, 
\label{eq:Y_n_r_n_T}
\end{align}
in which  $[\mathbf{u}_{n_T}(\theta_{k,n_{\mathrm{r}}}, \phi_k)]_t=\mathbf{a}(\theta_{k,n_{\mathrm{r}}})^T\bOmega_{k}(t,n_T)\mathbf{a}(\phi_k)$, where $\bOmega_{k}(t, n_T)=\bPhi_k((n_T-1)T+t)$ for $t\in \mathcal{T}\triangleq\{1,\dots, T\}$, $\mathbf{d}_\mathrm{d}(f_{_{\mathrm{d},k,n_{\mathrm{r}}}})=[ 1 \ e^{j 2\pi f_{_{\mathrm{d},k,n_{\mathrm{r}}}} T_{\mathrm{d}}} \dots e^{j 2\pi f_{_{\mathrm{d},k,n_{\mathrm{r}}}} T_{\mathrm{d}} (T-1)}]^T$,
and 
 $\mathbf{W}_{n_{\mathrm{r}},n_T}$ is the \ac{AWGN} matrix with variance $\sigma^2_{\mathbf{W}}=\Delta_f N_0$ in which $N_0$ is the power spectral density.

We design the phase profile matrix of the $k^{\th}$  \ac{RIS} as $\bOmega_{k}(t, n_T)=\omega_{k,t}\bOmega_{k}^{(n_T)}$ for $t\in \mathcal{T}$ and $n_T\in \mathcal{N}_T$, where matrix $\bOmega_{k}^{(n_T)}$ is constant  during each $T$ transmissions.
  ~We set vectors $\bomega_k=[\omega_{k,1}, \dots, \omega_{k,T}]^T$, $k\in \mathcal{K}$,  such that   $\forall \ t\in \{1,...,T/2\}$, $\omega_{k,2t}=-\omega_{k,2t-1}$ and  vectors $\bgamma_k=[\omega_{k,1}, \omega_{k,3},  \dots, \omega_{k,T-1}]^T$, for  $ k\in\mathcal{K}$  be orthogonal to each other. 
  To satisfy the orthogonality,  vectors $\bgamma_k$  can be chosen from the set of the columns of the \ac{FFT} matrix with dimensions $T/2$.  In practice, to not to be have to update the dimensions of the \ac{FFT} matrix when the number of the \acp{RIS} increases, we consider a maximum number for the RISs and set the dimensions of the \ac{FFT} matrix by this number. 
  
  Now, let us separate matrix $\mathbf{Y}_{n_{\mathrm{r}}, n_T}$ into sub-matrices $\mathbf{Y}_{n_{\mathrm{r}}, n_T,o}$ and $\mathbf{Y}_{n_{\mathrm{r}}, n_T,e}$ that  contain the odd and even columns, respectively.  Considering that the scatterers are  fixed during transmissions and thus the terms of the scatterers are the same in the odd and even columns, while by designing  $\omega_{k,2t}=-\omega_{k,2t-1}$ for $ t \in\{1,...,T/2\}$, the terms corresponding to the \acp{RIS} are opposite, we can remove the contributions of the scatterers by computing 
  \begin{equation} \mathbf{Y}_{n_{\mathrm{r}}, n_T}^{\prime}=\frac{1}{2}(\mathbf{Y}_{n_{\mathrm{r}}, n_T,o}-\mathbf{Y}_{n_{\mathrm{r}}, n_T,e}).
  \end{equation}

  \subsection{Joint Estimation of $\tau_{k,n_{\mathrm{r}}}$ and $f_{_{\mathrm{d},k,n_{\mathrm{r}}}}$ }
  To estimate $\tau_{k,n_{\mathrm{r}}}$ and $f_{_{\mathrm{d},k,n_{\mathrm{r}}}}$, we first extract the signal of the $k^{\th}$ \ac{RIS}.
  Neglecting the effect of the Doppler shifts, considering the orthogonality of $\bgamma_k$s, we extract the $n_T^{\th}$ signal vector received through the path including the $k^{\th}$ \ac{RIS}, by computing $\mathbf{r}_{k, n_{\mathrm{r}}, n_T}=\frac{2}{T}\mathbf{Y}_{n_{\mathrm{r}}, n_T}^{\prime}\bgamma_{k}^{*}$. Thus,
  
\begin{align}
&\mathbf{r}_{k,n_{\mathrm{r}}, n_T}= \sqrt{E}_s \  g_{k, n_{\mathrm{r}}} \mathbf{d}(\tau_{k,n_{\mathrm{r}}}) \mathbf{a}(\theta_{k,n_{\mathrm{r}}})^T \bOmega_{k}^{(n_T)}\mathbf{a}(\phi_k ) \times \notag \\ & e^{j 2\pi f_{_{\mathrm{d},k,n_{\mathrm{r}}}} (n_T-1) T_{\mathrm{d}} T}\sum_{t=1}^{T}e^{j 2\pi f_{_{\mathrm{d},k,n_{\mathrm{r}}}} T_{\mathrm{d}} (t-1)}  +\mathbf{z}_{k, n_{\mathrm{r}}, n_T},\label{eq:r_k_n_T}
\end{align}
 in which $\mathbf{z}_{k,n_{\mathrm{r}},n_T}=\frac{2}{T}\mathbf{W}_{n_{\mathrm{r}},n_T}^{\prime}\bgamma_{k}^{*}+\mathcal{I}_{k,n_{\mathrm{r}},n_T}$, where $\mathcal{I}_{k,n_{\mathrm{r}},n_T}$ is the leakage of the signals corresponding to the other RISs. 
We must note that neglecting the Doppler shifts to be able to exactly extract the signal reflected by each \ac{RIS} as \eqref{eq:r_k_n_T}, $T$ should satisfy $K\leq T/2$.

Now, considering the design of matrix $\bOmega_{k}^{(n_T)}$ in which for $n_T\in\{1,\dots, N_T/2\}$ and $m\neq 2$, $\Omega_{k,m,m}^{(2n_T-1)}=\Omega_{k,m,m}^{(2n_T)}$  and $\Omega_{k,2,2}^{(2n_T-1)}\neq\Omega_{k,2,2}^{(2n_T)}$ (See Section \textrm{III}-\textit{C}), we subtract the consecutive vectors i.e., $\mathbf{r}_{k,n_{\mathrm{r}}, 2n_T-1}$ and $\mathbf{r}_{k,n_{\mathrm{r}}, 2n_T}$ for $n_T\in\{1,\dots, N_T/2\}$ to obtain a set of vectors that their phase shifts compared to each other  is just the Doppler shift as blow
\begin{equation}\mathbf{r}^{\prime}_{k,n_{\mathrm{r}}, n_T}=\mathbf{r}_{k,n_{\mathrm{r}}, 2n_T-1}-\mathbf{r}_{k,n_{\mathrm{r}}, 2n_T}, \forall n_T\in\{1,\dots, N_T/2\}. \label{eq:r_prime}
\end{equation}
 Then, we construct matrix $\mathbf{R}_{k,n_{\mathrm{r}}}=[\mathbf{r}^{\prime}_{k,n_{\mathrm{r}}, 1} \dots \mathbf{r}^{\prime}_{k,n_{\mathrm{r}}, N_T/2}] \in \mathcal{C}^{N\times N_T/2}$ in which the difference of the phase shifts of the consecutive elements on each  column  is $2\pi \Delta_f \tau_{k,n_{\mathrm{r}}}$ and the  difference of the phase shifts the consecutive elements on each  row is equal to $4\pi  f_{_{\mathrm{d},k,n_{\mathrm{r}}}}T_{\mathrm{d}} T$. 
We aim to employ 2DFFT of signal $\mathbf{r}_{k,n_{\mathrm{r}}, n_T}$ to jointly estimate $\tau{k,n_{\mathrm{r}}}$ and $f_{_{\mathrm{d},k,n_{\mathrm{r}}}}$. But, the challenge is that the sign  of $f_{_{\mathrm{d},k,n_{\mathrm{r}}}}$ is unknown\footnote{Because the value of $f_{_{\mathrm{d},k,n_{\mathrm{r}}}}$ depends on the position of the \ac{RIS} and depending on it $f_{_{\mathrm{d},k,n_{\mathrm{r}}}}$ can get a positive or a negative value.}. To solve this problem, assuming that 
$|4\pi  f_{_{\mathrm{d},k,n_{\mathrm{r}}}}T_{\mathrm{d}} T|<2 \pi-|4\pi  f_{_{\mathrm{d},k,n_{\mathrm{r}}}}T_{\mathrm{d}} T|$, we propose algorithm 1 to jointly estimate $\tau_{k,n_{\mathrm{r}}}$ and $f_{_{\mathrm{d},k,n_{\mathrm{r}}}}$. In this algorithm, $\mathbf{F}_{\tau} \in \mathcal{C}^{N_{f, \tau}\times N}$ and $\mathbf{F}_{f_{_{\mathrm{d}}},1} \in \mathcal{C}^{N_{f,f_{_{\mathrm{d}}}}\times N_T/2}$ are FFT matrices, and $\mathbf{F}_{f_{_{\mathrm{d}}},2} \in \mathcal{C}^{N_{f, f_{_{\mathrm{d}}}}\times N_T/2}$ is an IFFT matrix. In this algorithm, we use this subject that if we one time estimate $f_{_{\mathrm{d},k,n_{\mathrm{r}}}}$ employing the FFT matrix and one time with IFFT matrix, the estimated values satisfy $$|4\pi\hat{f}_{_{\mathrm{d},k,n_{\mathrm{r}},1}}T_{\mathrm{d}} T|=2\pi-|4\pi \hat{f}_{_{\mathrm{d},k,n_{\mathrm{r}},2}} T_{\mathrm{d}} T|.$$ 

\begin{algorithm}[t]
\caption{Proposed Doppler frequency and ToA Estimation Scheme}
\label{alg_f_tau}
\begin{algorithmic}[1]
\FOR{$n_{\mathrm{r}}=1$ to $N_{\mathrm{r}}$}
\FOR{$k=1$ to $K$}
\STATE Set
$$[n_{1,\tau}^{*},n_{1,f_{_{\mathrm{d}}}}^{*}]= \arg \max \left| \left[ \mathbf{F}_{\tau}\mathbf{R}_{k,n_{\mathrm{r}}}\mathbf{F}_{f_{_{\mathrm{d}}},1}^T\right]_{n_{1,\tau},n_{1,f_{_{\mathrm{d}}}}} \right|.$$
\STATE Compute $\hat{\tau}_{k,n_{\mathrm{r}}}=(n_{1,\tau}^{*}-1)/(N_{F_{\tau}}\Delta_f) $ and $\hat{f}_{_{\mathrm{d},k,n_{\mathrm{r}},1}}=(n_{1,f_{_{\mathrm{d}}}}^{*}-1)c/(2 T T_{\mathrm{d}} f_c)$.
\STATE Set
$$[n_{2,\tau}^{*},n_{2,f_{_{\mathrm{d}}}}^{*}]= \arg \max \left| \left[ \mathbf{F}_{\tau}\mathbf{R}_{k,n_{\mathrm{r}}}\mathbf{F}_{f_{_{\mathrm{d}}},2}^T\right]_{n_{2,\tau},n_{1,f_{_{\mathrm{d}}}}} \right|. $$
\STATE Compute $\hat{f}_{_{\mathrm{d},k,n_{\mathrm{r}},2}}=-(n_{2,f_{_{\mathrm{d}}}}^{*}-1)c/(2 T T_{\mathrm{d}} f_c)$.
\IF{$|\hat{f}_{_{\mathrm{d},k,n_{\mathrm{r}},1}}|<|\hat{f}_{_{\mathrm{d},k,n_{\mathrm{r}},2}}|$}
\STATE $\hat{f}_{_{\mathrm{d},k,n_{\mathrm{r}}}}=\hat{f}_{_{\mathrm{d},k,n_{\mathrm{r}},1}}$.
\ELSE
\STATE$\hat{f}_{_{\mathrm{d},k,n_{\mathrm{r}}}}=\hat{f}_{_{\mathrm{d},k,n_{\mathrm{r}},2}}$.
\ENDIF
 \ENDFOR
\ENDFOR
\end{algorithmic}
\end{algorithm} 
\subsection{Estimation of $\alpha_{k,n_{\mathrm{r}}}$}
In order to estimate $\alpha_{k,n_{\mathrm{r}}}$, we first remove the effect of ${f}_{d,k,n_{\mathrm{r}}}$ and  $\tau_{k,n_{\mathrm{r}}}$.
To eliminate the effect of Doppler shifts, for each  $k \in \mathcal{K}$, $n_T \in  \mathcal{N}_T $ and $n_{\mathrm{r}} \in \mathcal{N}_{\mathrm{r}}$, we construct  semi-orthogonal vectors $\bgamma_{k,n_{\mathrm{r}}, n_T}$,  by pointwise multiplication of vectors $\bgamma_k$s and vectors 
\begin{align*}
& e^{j2\pi \hat{f}_{_{\mathrm{d},k,n_{\mathrm{r}}}}T_{\mathrm{d}} T (n_T-1)}\times \left[(1\hspace{-0.5mm}+\hspace{-0.5mm}e^{j2\pi \hat{f}_{_{\mathrm{d},k,n_{\mathrm{r}}}}T_{\mathrm{d}} }), \dots, \right. \\  & \left. (e^{j2\pi \hat{f}_{_{\mathrm{d},k,n_{\mathrm{r}}}}T_{\mathrm{d}} (T-2)}\hspace{-0.5mm}+\hspace{-0.5mm}e^{j2\pi \hat{f}_{_{\mathrm{d},k,n_{\mathrm{r}}}}T_{\mathrm{d}}  (T-1)})\right]^T\hspace{-3mm},
\end{align*}
 for $k\in \mathcal{K}$ \footnote{The vectors will be orthogonal when the length of them tends to infinity.}. Then,  we simultaneously extract the signal of the $k^{\th}$ RIS and remove the effect of the Doppler shift and $\tau_{k,n_{\mathrm{r}}}$
 by obtaining
\begin{align}\label{eq:remove_tau}
\mathrm{r}_{k,n_{\mathrm{r}},n_T}^{\prime\prime}&=\frac{2}{T}(\mathbf{Y}_{n_{\mathrm{r}}, n_T}^{\prime}\bgamma_{k,n_{\mathrm{r}}, n_T})\odot \mathbf{d}(\hat{\tau}_{k,n_{\mathrm{r}}}) \\& \simeq s_{k, n_{\mathrm{r}}, n_T} 1_{N}+\mathbf{z}_{k,n_{\mathrm{r}}, n_T}^{\prime}, \notag
\end{align}
in which  $s_{k, n_{\mathrm{r}}, n_T}\triangleq  \sum_{m=1}^{M} \sqrt{E}_s g_{k, n_{\mathrm{r}}} \Omega_{k,m,m}^{(n_T)} e^{j \frac{2\pi}{\lambda} (m-1)d  \alpha_{k,n_{\mathrm{r}}}}$ and $\mathbf{z}_{k,n_{\mathrm{r}}, n_T}^{\prime}$ the AWGN vector.

Regarding  invariance property of the \ac{ML} estimation approach\footnote{The invariance property states that if $\hat{\theta}$ is the \ac{ML} estimation of the $\theta$, then for function $f$ the \ac{ML} estimation of $f(\theta)$ is $f(\hat{\theta})$ \cite{kay1993fundamentals}.}, we first estimate  $s_{k, n_{\mathrm{r}}, n_T}$, and then use it to obtain the estimated value of $\alpha_{k,n_{\mathrm{r}}}$. The \ac{ML} estimation  of $s_{k, n_{\mathrm{r}}, n_T}$ for $n_T\in \mathcal{N}_T$ and $n_{\mathrm{r}} \in \mathcal{N}_{\mathrm{r}}$ is 
\begin{equation}\label{eq:s_hat}
\hat{s}_{k,  n_{\mathrm{r}}, n_T}=\frac{1}{N} \sum_{n=1}^{N} [\mathbf{r}_{k, n_{\mathrm{r}},  n_T}^{\prime\prime}]_n.
\end{equation}
Now, for $n_{\mathrm{r}}\in \mathcal{N}_{\mathrm{r}}$ we can obtain   $\hat{\alpha}_{k, n_{\mathrm{r}}}$, by solving the following system of non-linear equations
\begin{align}
&\sum_{m=1}^{M} \sqrt{E}_s g_{k, n_{\mathrm{r}}} \Omega_{k,m,m}^{(N_T)} e^{j \frac{2\pi}{\lambda} (m-1)d  \alpha_{k,n_{\mathrm{r}}}}\hspace{-1mm}=\hspace{-0.5mm}\hat{s}_{k, n_{\mathrm{r}}, n_T},  \forall n_T \in \mathcal{N}_T.
\label{eq:alpha_k_gr_1} 
\end{align}

Regarding the degree of each equation in terms of $e^{j \frac{2\pi}{\lambda}d  \alpha_{k,n_{\mathrm{r}}}}$, there exist up to $M-1$ distinct values for $\alpha_{k,n_{\mathrm{r}}}$ in each interval of length  $\frac{\lambda}{d}$ satisfying this equation system. Consequently,
  we generally have ambiguity to estimate the correct value of $\alpha_{k,n_{\mathrm{r}}}$. To address this challenge, we design $\Omega_{k,m,m}^{(n_T)}$ for $n_T\in \mathcal{N}_T$ and $m\in\{2,\dots,M\}$ and  employ the elimination approach  such that the correct value of  $\alpha_{k,n_{\mathrm{r}}}$ is distinguishable. For simplicity, we perform this task in such a way that only one of the powers of $e^{j \frac{2\pi}{\lambda}d  \alpha_{k,n_{\mathrm{r}}}}$ remains. 

{ Considering that the RISs can have orientation generally  $-2<\alpha_{k,n_{\mathrm{r}}}<2$. Thus, assuming that $d=\frac{\lambda}{L}$,  the value of $\frac{2 \pi }{\lambda} (m-1) d \alpha_{k,n_{\mathrm{r}}}$ is in the range $[-(m-1)\frac{\pi}{L},(m-1)\frac{\pi}{L}] $.
Therefore,  the remained exponent of $e^{j \frac{2\pi}{\lambda} d  \alpha_{k,n_{\mathrm{r}}}}$ ($M^{\prime}$) should  be equal or less than  $L$, where $L\geq 4$  to ensure that $\frac{2 \pi }{\lambda} M^{\prime} d \alpha_{k,n_{\mathrm{r}}}$ is in $(-\pi,\pi)$ and thus the correct value of $\alpha_{k,n_{\mathrm{r}}}$ is distinguishable}\footnote{ Actually, under this condition we  ensure that in an equation such as $e^{j \frac{2\pi}{M}^{\prime}{\lambda} d  \alpha_{k,n_{\mathrm{r}}}}=c_0$ the value of $\alpha_{k,n_{\mathrm{r}}}$, is equal to $\frac{\lambda}{{M}^{\prime}d}\arg\{c_0\},$ and do not have ambiguity between  values $\frac{\lambda}{{M}^{\prime}d}\arg\{c_0\}+2k \pi /{M}^{\prime}$. }. 
  To ensure that our system works for all values of $L$ that satisfy this constraint,  we design the phase shifts such that for $n_T\in\{1,\dots, N_T/2\}$ and $m\neq 2$, $\Omega_{k,m,m}^{(2n_T-1)}=\Omega_{k,m,m}^{(2n_T)}$  and $\Omega_{k,2,2}^{(2n_T-1)}\neq\Omega_{k,2,2}^{(2n_T)}$. Thus, subtracting the consecutive equations in \eqref{eq:alpha_k_gr_1}, we obtain the following  system of equations 
\begin{align}
& \notag g_{k,n_{\mathrm{r}}} e^{j \frac{2\pi  d}{\lambda} \alpha_{k,n_{\mathrm{r}}} }=\frac{1}{\Omega_{k,2,2}^{(N_T-1)}-\Omega_{k,2,2}^{(N_T)}}(\hat{s}_{k,n_{\mathrm{r}},n_T-1}-\hat{s}_{k,n_{\mathrm{r}},n_T}), \\ &  \ n_T\in\{2,\dots,N_T\}.
\end{align}
Then, we formulate the following \ac{LS} problem to estimate $\alpha_{k,n_{\mathrm{r}}}$
\begin{equation}
\min_{\alpha_{k,n_{\mathrm{r}}} \in \mathbb{R}} \sum_{n_T=1}^{N_T/2} \left|g_{k,n_{\mathrm{r}}} e^{j \frac{2\pi  d}{\lambda} \alpha_{k,n_{\mathrm{r}}} }-\frac{\hat{s}_{k, n_{\mathrm{r}}, 2n_T-1}-\hat{s}_{k,n_{\mathrm{r}},2n_T}}{\Omega_{k,2,2}^{(2n_T-1)}-\Omega_{k,2,2}^{(2n_T)}}\right|^2, \label{eq:LS_alpha}
\end{equation}
and thus, the optimal value of $\alpha_{k,n_{\mathrm{r}}}$ can be derived as
\begin{equation}
\label{alpha_hat}
\hat{ \alpha}_{k,n_{\mathrm{r}}}=\frac{\lambda}{2 \pi d}\arg\left(\frac{1}{N_T} \sum_{n_T=1}^{N_T/2}\frac{\hat{s}_{k, n_{\mathrm{r}}, 2n_T-1}-\hat{s}_{k,n_{\mathrm{r}},2n_T}}{\Omega_{k,2,2}^{(2n_T-1)}-\Omega_{k,2,2}^{(2n_T)}}\right).
\end{equation}
\begin{algorithm}[t]
\caption{Proposed \ac{RIS} Localization Approach}
\label{alg1}
\begin{algorithmic}[1]
\STATE The transmitter sends the signal for $\bar{T}$ times. 
\FOR{$n_{\mathrm{r}}=1$ to $N_{\mathrm{r}}$}
\FOR{$n_T=1$ to $N_T$}
\STATE Construct matrix $\mathbf{Y}_{n_{\mathrm{r}}, n_T}$ and calculate $ \mathbf{Y}_{n_{\mathrm{r}}, n_T}^{\prime}=\frac{1}{2}(\mathbf{Y}_{n_{\mathrm{r}}, n_T,o}-\mathbf{Y}_{n_{\mathrm{r}}, n_T,e})$ to remove the effect of the scatters.
\FOR{$k=1$ to $K$}
\STATE Extract the signal reflected by the $k^{\th}$ \ac{RIS} by computing $\mathbf{r}_{k, n_{\mathrm{r}}, n_T}=\frac{2}{T}\mathbf{Y}_{n_{\mathrm{r}}, n_T}^{\prime}\bgamma_{k}^{*}$.
\ENDFOR
\ENDFOR
\FOR{$k=1$ to $K$}
\FOR {$n_T=1$ to $N_T$}
\STATE Obtain signal $\mathbf{r}_{k, n_{\mathrm{r}}, n_T}^{\prime}$ using \ref{eq:r_prime}.
\ENDFOR
\STATE Estimate ${\tau}_{k,n_{\mathrm{r}}}$ and $f_{_{\mathrm{d},k,n_{\mathrm{r}}}}$ using algorithm \ref{alg_f_tau}.
\STATE Remove the effect of ${\tau}_{k,n_{\mathrm{r}}}$ using \eqref{eq:remove_tau}.
\FOR{$n_T=1$ to $N_T$}
\STATE Estimate ${s}_{k,n_{\mathrm{r}},n_T}$ using \eqref{eq:s_hat}.
\ENDFOR
 \STATE Estimate ${\alpha}_{k,n_{\mathrm{r}}}$ using \eqref{alpha_hat}.
 \ENDFOR
\ENDFOR
\FOR{$k=1$ to $K$}
\STATE Define problem \eqref{eq:opt_localization} by substituting $\hat{{\tau}}_{k,n_{\mathrm{r}}}$ and $\hat{{\alpha}}_{k,n_{\mathrm{r}}}$ and solve it  to estimate the location of the $k^{\th}$ \ac{RIS}.
\ENDFOR
\end{algorithmic}
\end{algorithm} 
\begin{algorithm}[htpb]
\caption{Proposed \ac{EKF}-based \ac{RIS} Tracking Approach }
\label{alg2}
\small
\begin{algorithmic}[1]
\STATE \textbf{Initialization:} For $k\in \mathcal{K}$, estimate ${\bf x}_k[0|0]$ employing algorithm \ref{alg1}. Set $n=1$.
\WHILE{Tracking the \acp{RIS}}
\FOR{$k=1$ to $K$}
\STATE Predict the location and  velocity of the $k^{\th}$ \ac{RIS} as follows:
\begin{equation*}
{\bf x}_k[n|n-1]={\bf A}~ {\bf x}_k[n-1|n-1]+{\bf B}~{\bf a}_k.
\end{equation*}
\STATE  
Obtain the minimum prediction \ac{MSE} matrix: 
\begin{equation*}
{\bf M}_k[n|n-1]={\bf A}{\bf M}_k[n-1|n-1]{\bf A}^{T}+ {\bf Q}_k[n].
\end{equation*}
\STATE Construct vector ${\bf h}({\bf x}_k[n])$ as
\begin{equation*}
{\bf h}({\bf x}_k[n])=
\scriptsize{\begin{bmatrix}
\left\|\pk[n]-\pt\right\|+\left\|\pk[n]-\prf\right\|\\ 
\vdots
\\ 
\left\|\pk[n]-\pt\right\|+\left\|\pk[n]-\prNr\right\|
\\ \\
\cos \psi_{k,1} \left(\frac{\left|p_{_{k,1}}-p_{_{\mathrm{t},1}}\right|}{\left\|{\mathbf{p}}_{_k}-{\mathbf{p}}_{_{\mathrm{t}}}\right\|}- \frac{\left|p_{_{k,1}}- {p}_{_{{\mathrm{r}}, 1,1}}\right|}{\left\|{\mathbf{p}}_{_{k}}-{\mathbf{p}}_{_{{\mathrm{r}},1}}\right\|}\right)+\\+\sin \psi_{k,1} \left(\frac{\left|p_{_{k,2}}-p_{_{\mathrm{t},2}}\right|}{\left\|{\mathbf{p}}_{_k}-{\mathbf{p}}_{_{\mathrm{t}}}\right\|}+ \frac{\left|p_{_{k,2}}- {p}_{_{{\mathrm{r}}, 1,2}}\right|}{\left\|{\mathbf{p}}_{_{k}}-{\mathbf{p}}_{_{{\mathrm{r}},1}}\right\|}\right)
\\
\vdots
\\
\cos \psi_{k,N_{\mathrm{r}}} \left(\frac{\left|p_{_{k,1}}-p_{_{\mathrm{t},1}}\right|}{\left\|{\mathbf{p}}_{_k}-{\mathbf{p}}_{_{\mathrm{t}}}\right\|}- \frac{\left|p_{_{k,1}}- {p}_{_{{\mathrm{r}}, N_{\mathrm{r}},1}}\right|}{\left\|{\mathbf{p}}_{_{k}}-{\mathbf{p}}_{_{{\mathrm{r}},N_{\mathrm{r}}}}\right\|}\right)+\\+\sin \psi_{k,N_{\mathrm{r}}} \left(\frac{\left|p_{_{k,2}}-p_{_{\mathrm{t},2}}\right|}{\left\|{\mathbf{p}}_{_k}-{\mathbf{p}}_{_{\mathrm{t}}}\right\|}+ \frac{\left|p_{_{k,2}}- {p}_{_{{\mathrm{r}}, N_{\mathrm{r}},2}}\right|}{\left\|{\mathbf{p}}_{_{k}}-{\mathbf{p}}_{_{{\mathrm{r}},N_{\mathrm{r}}}}\right\|}\right)
\end{bmatrix}}
\end{equation*}
\STATE Obtain matrix ${\bf H}_k[n|n-1]$ as
\begin{equation*}
{\bf H}_k[n|n-1]=
\dfrac{\partial{\bf h}({\bf x}_k[n])}{\partial{\bf x}_k[n]}\Big|_{{\bf x}_k[n]={{\bf x}}_k[n|n-1]}.
\end{equation*}
\STATE
Calculate the Kalman gain matrix as
\begin{align*}
&{\bf K}_k[n]={\bf M}_k[n|n-1] {\bf H}_k^{H}[n|n-1] \\ & \times \left({\bf C}_k[n]+{\bf H}_k[n|n-1]{\bf M}_k[n|n-1]{\bf H}_k^{H}[n|n-1]\right)^{-1}. 
\end{align*}
\FOR{$n_{\mathrm{r}}=1$ to $N_{\mathrm{r}}$}
\STATE Estimate $\tau_{k,n_{\mathrm{r}}}$ using algorithm 1.
\STATE Estimate $\alpha_{k,n_{\mathrm{r}}}$ as follows:
\begin{equation*}
\hspace{-4mm}\hat{ \alpha}_{k,n_{\mathrm{r}}}\hspace{-1mm}=\hspace{-1mm}\frac{\lambda}{2 \pi d}\arg \hspace{-1mm}\left(\hspace{-1mm}\frac{1}{N_T} \sum_{n_T=1}^{N_T/2}\frac{\hat{s}_{k, n_{\mathrm{r}}, 2n_T-1}-\hat{s}_{k,n_{\mathrm{r}},2n_T}}{\Omega_{k,2,2}^{(2n_T-1)}-\Omega_{k,2,2}^{(2n_T)}}\hspace{-1mm}\right).
\end{equation*}
\ENDFOR
\STATE Construct the measurement vector ${\bf \bnu}_k[n]$ using   $\hat{\tau}_{k,n_{\mathrm{r}}}$ and $\hat{\alpha}_{k,n_{\mathrm{r}}}$, $n_{\mathrm{r}}\in \mathcal{N}_{\mathrm{r}}$.
\STATE 
Correct the estimation of position and velocity of the \ac{RIS}  as
\begin{equation*}{\bf x}_k[n|n]\hspace{-0.5mm}=\hspace{-0.5mm}{\bf x}_k[n|n-1]+{\bf K}_k[n] \hspace{-0.5mm}\left({\bf \bnu}_k[n]\hspace{-0.5mm}-\hspace{-0.5mm}{\bf h}(\hat{\bf x}_k[n|n-1])\right).
\end{equation*}
\STATE Obtain the minimum \ac{MSE} matrix  
\begin{equation*}
{\bf M}_k[n|n]=\left({\bf I} -{\bf K}_k[n]{\bf H}_k[n]\right){{\bf M}_k[n|n-1]}. 
\end{equation*}
\ENDFOR
\STATE Set $n=n+1$.
\ENDWHILE
\end{algorithmic}
\end{algorithm}
\subsection{The Localization Approach}
Now, we aim to define an \ac{LS} problem considering the estimated values of $\tau_{k,n_{\mathrm{r}}}$ and $\alpha_{k,n_{\mathrm{r}}}$ for $n_{\mathrm{r}}\in \mathcal{N}_{\mathrm{r}}$, to localize the $k^{\th}$ \ac{RIS}. In this regard, we first derive the equation of $\tau_{k,n_{\mathrm{r}}}$ and $\alpha_{k,n_{\mathrm{r}}}$ in terms of the location of the \ac{RIS}. 

Defining $c$ as the speed of light, we have
\begin{equation}
\label{eq:tau_hat}
\left\|\pk-\pt\right\|+\left\|\pk-\pr\right\|=c\tau_{k,n_{\mathrm{r}}}.
\end{equation}

 Also, defining $\phi_{k,n_{\mathrm{r}}}^{\prime}$ and $\theta_{k,n_{\mathrm{r}}}^{\prime}$ as shown in Fig. \ref{parallel_scenario}, we have 
 $\phi_{k,n_{\mathrm{r}}}=\phi_{k,n_{\mathrm{r}}}^{\prime}-\psi_{k,n_{\mathrm{r}}}$ and $\theta_{k,n_{\mathrm{r}}}=\theta_{k,n_{\mathrm{r}}}^{\prime}-\psi_{k,n_{\mathrm{r}}}$, and using it  the value of $\alpha_{k,n_{\mathrm{r}}}$ would be equal to
\begin{align}
\alpha_{k,n_{\mathrm{r}}}=&\cos \psi_{k,n_{\mathrm{r}}} \left(\frac{\left|p_{_{k,1}}-p_{_{\mathrm{t},1}}\right|}{\left\|{\mathbf{p}}_{_k}-{\mathbf{p}}_{_{\mathrm{t}}}\right\|}- \frac{\left|p_{_{k,1}}- {p}_{_{{\mathrm{r}}, n_{\mathrm{r}},1}}\right|}{\left\|{\mathbf{p}}_{_{k}}-{\mathbf{p}}_{_{{\mathrm{r}},n_{\mathrm{r}}}}\right\|}\right)+ \notag \\
&\sin \psi_{k,n_{\mathrm{r}}} \left(\frac{\left|p_{_{k,2}}-p_{_{\mathrm{t},2}}\right|}{\left\|{\mathbf{p}}_{_k}-{\mathbf{p}}_{_{\mathrm{t}}}\right\|}+ \frac{\left|p_{_{k,2}}- {p}_{_{{\mathrm{r}}, n_{\mathrm{r}},2}}\right|}{\left\|{\mathbf{p}}_{_{k}}-{\mathbf{p}}_{_{{\mathrm{r}},n_{\mathrm{r}}}}\right\|}\right),\label{eq:alpha_hat}
\end{align}
where $p_{k,1}=[\pk]_1$, $p_{k,2}=[\pk]_2$, $\ptOne=[\pt]_1$, $\ptTwo=[\pt]_2$, $\prOne=[\pr]_1$, and $\prTwo=[\pr]_2$.
Now, considering the estimated values of $\tau_{k,n_{\mathrm{r}}}$ and $\alpha_{k,n_{\mathrm{r}}}$ and equations \eqref{eq:tau_hat} and \eqref{eq:alpha_hat}, we define the following \ac{LS} problem to localize the $k^{\th}$ \ac{RIS}
\begin{align}
\min_{\pk} &\sum_{n_{\mathrm{r}}=1}^{N_{\mathrm{r}}}\Big( \left| \hspace{0.05cm} \left\| {\mathbf{p}}_{_k}-{\mathbf{p}}_{_{\mathrm{t}}}\right\|+\left\| {\mathbf{p}}_{_k}-{\mathbf{p}}_{_{{\mathrm{r}},n_{\mathrm{r}}}}\right\|-c\hat{\tau}_{k,n_{\mathrm{r}}}\right|^2+\notag  \\ 
&  \left|\cos \psi_{k,n_{\mathrm{r}}} \left(\frac{\left|p_{_{k,1}}-p_{_{\mathrm{t},1}}\right|}{\left\|{\mathbf{p}}_{_k}-{\mathbf{p}}_{_{\mathrm{t}}}\right\|}- \frac{\left|p_{_{k,1}}- {p}_{_{{\mathrm{r}}, n_{\mathrm{r}},1}}\right|}{\left\|{\mathbf{p}}_{_{k}}-{\mathbf{p}}_{_{{\mathrm{r}},n_{\mathrm{r}}}}\right\|}\right)+\label{eq:opt_localization} \right. \\ & \left.
\sin \psi_{k,n_{\mathrm{r}}} \left(\frac{\left|p_{_{k,2}}-p_{_{\mathrm{t},2}}\right|}{\left\|{\mathbf{p}}_{_k}-{\mathbf{p}}_{_{\mathrm{t}}}\right\|}+ \frac{\left|p_{_{k,2}}- {p}_{_{{\mathrm{r}}, n_{\mathrm{r}},2}}\right|}{\left\|{\mathbf{p}}_{_{k}}-{\mathbf{p}}_{_{{\mathrm{r}},n_{\mathrm{r}}}}\right\|}\right)   -\hat{\alpha}_{k,n_{\mathrm{r}}}\right|^{2}\Big),  \notag
\end{align}
which  can be solved by quasi-Newton approaches.

  Algorithm \ref{alg1} summarizes  the proposed localization approach.
   Considering the number of multiplications in  extracting signal $\mathbf{r}_{k, n_{\mathrm{r}}, n_T}$, the number of multiplications  in algorithm 2, the number of multiplications in obtaining signal $\mathbf{r}_{k, n_{\mathrm{r}}, n_T}^{\prime\prime}$ and the order of complexity of finding $n_{f,f_{_{\mathrm{d}}}}$ and $n_{f, \tau}$ in algorithm 1,
  the computational complexity of this localization approach is of the order of $\mathcal{O}\left(KN_{\mathrm{r}}(NN_T\hspace{-0.5mm}+\hspace{-0.5mm}N_T N_{f,\tau} N_{f,f_{_{\mathrm{d}}}}\hspace{-1mm}+\hspace{-1mm}N_{f,\tau} N_{f,f_{_{\mathrm{d}}}}\hspace{-1mm} \log(N_{f,\tau} N_{f,f_{_{\mathrm{d}}}})\right)$.

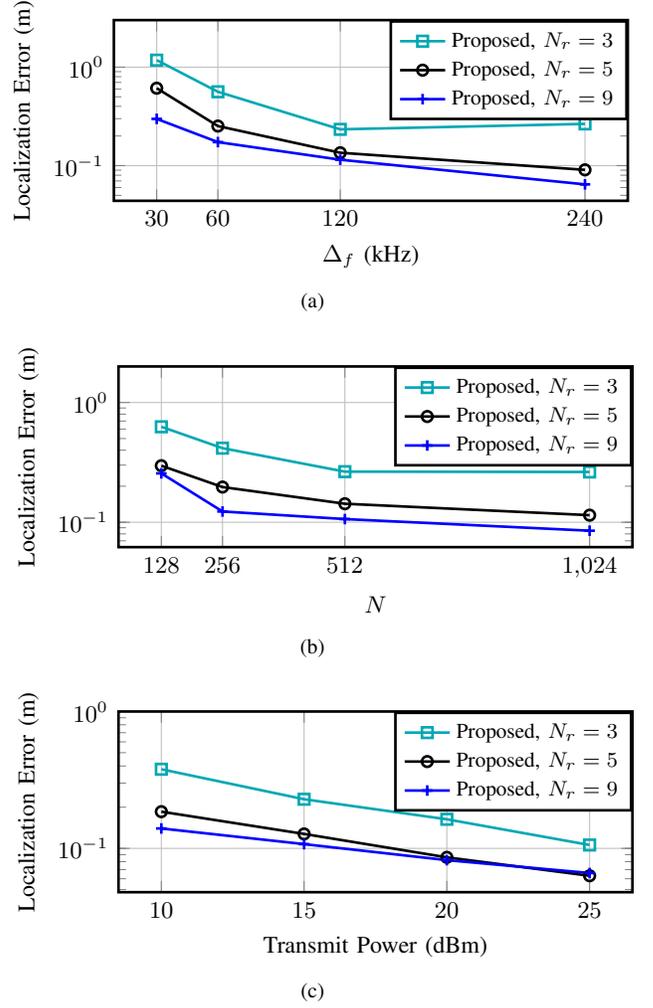
\begin{figure}[t] 
\begin{center}
	\subfigure[]{\pgfplotsset{every axis/.append style={
		font=\small,
		line width=1pt,
			legend style={font=\footnotesize,  at={(1,1)},anchor=north east},legend cell align=left},
} %
\pgfplotsset{compat=1.13}
	\begin{tikzpicture}
\begin{axis}[
xlabel near ticks,
ylabel near ticks,
ymode=log,
grid=major,
xlabel={$\Delta_f$ (kHz) },
ylabel={Localization Error (m) },
	xtick={30, 60,120,240},
yticklabel style={/pgf/number format},
width=0.95\linewidth,
height=0.45\linewidth,
	legend columns=1,	
legend entries={
{Proposed,  $N_r=3$},
{Proposed, $N_r=5$},
 {Proposed,  $N_r=9$}},
	 ymax=3,
ylabel style={font=\small},
xlabel style={font=\small},
]
\definecolor{green-samaneh}{rgb}{0, 0.66, 0.7}
\addplot[green-samaneh,solid,mark=square] table {Figures/Fig_loc_error_vs_delta_f_diff_Nr/loc_error_vs_delta_f_Psi_pi_6_T_16_N_T_8_N_512_6GHz_d_4_1.dat};
\addplot[black,solid,mark=o] table {Figures/Fig_loc_error_vs_delta_f_diff_Nr/loc_error_vs_delta_f_Psi_pi_6_T_16_N_T_8_N_512_6GHz_d_4_Nr_5.dat};
\addplot[blue,solid,mark=+] table {Figures/Fig_loc_error_vs_delta_f_diff_Nr/loc_error_vs_delta_f_Psi_pi_6_T_16_N_T_8_N_512_6GHz_d_4_Nr_9.dat};


\end{axis}
\end{tikzpicture} \label{fig:Nr_delta_f}}
	\subfigure[]{\pgfplotsset{every axis/.append style={
		font=\small,
		line width=1pt,
			legend style={font=\footnotesize,  at={(1,1)},anchor=north east},legend cell align=left},
} %
\pgfplotsset{compat=1.13}
	\begin{tikzpicture}
\begin{axis}[
xlabel near ticks,
ylabel near ticks,
ymode=log,
grid=major,
xlabel={$N$ },
ylabel={Localization Error (m) },
	xtick={128, 256,512,1024,2048},
yticklabel style={/pgf/number format},
width=0.95\linewidth,
height=0.45\linewidth,
	legend columns=1,	
legend entries={
{Proposed,  $N_r=3$},
{Proposed, $N_r=5$},
 {Proposed,  $N_r=9$}},
	ymax=2,
ylabel style={font=\small},
xlabel style={font=\small},
]
\definecolor{green-samaneh}{rgb}{0, 0.66, 0.7}
\addplot[green-samaneh,solid,mark=square] table {Figures/Fig_loc_error_vs_N_diff_Nr/loc_error_vs_N_Psi_pi_6_T_16_N_T_8_delta_f_120_6GHz_d_4.dat};
\addplot[black,solid,mark=o] table {Figures/Fig_loc_error_vs_N_diff_Nr/loc_error_vs_N_Psi_pi_6_T_16_N_T_8_delta_f_120_6GHz_d_4_Nr_5.dat};
\addplot[blue,solid,mark=+] table {Figures/Fig_loc_error_vs_N_diff_Nr/loc_error_vs_N_Psi_pi_6_T_16_N_T_8_delta_f_120_6GHz_d_4_Nr_9.dat};


\end{axis}
\end{tikzpicture}\label{fig:Nr_N}}
	\subfigure[]{\pgfplotsset{every axis/.append style={
		font=\small,
		line width=1pt,
			legend style={font=\footnotesize,  at={(1,1)},anchor=north east},legend cell align=left},
} %
\pgfplotsset{compat=1.13}
	\begin{tikzpicture}
\begin{axis}[
xlabel near ticks,
ylabel near ticks,
ymode=log,
grid=major,
xlabel={Transmit Power (dBm) },
ylabel={Localization Error (m) },
yticklabel style={/pgf/number format},
width=0.95\linewidth,
height=0.45\linewidth,
legend entries={
{Proposed,  $N_r=3$},
{Proposed, $N_r=5$},
 {Proposed,  $N_r=9$}},
	ymin=0, 
	ymax=1,
ylabel style={font=\small},
xlabel style={font=\small},
]
\definecolor{green-samaneh}{rgb}{0, 0.66, 0.7}
\addplot[green-samaneh,solid,mark=square] table {Figures/Fig_loc_error_vs_P_diff_Nr/loc_error_vs_p_Psi_pi_6_T_16_N_T_8_delta_f_120_N_512_6GHz_d_4.dat};
\addplot[black,solid,mark=o] table {Figures/Fig_loc_error_vs_P_diff_Nr/loc_error_vs_p_Psi_pi_6_T_16_N_T_8_delta_f_120_N_512_6GHz_d_4_Nr_5.dat};
\addplot[blue,solid,mark=+] table {Figures/Fig_loc_error_vs_P_diff_Nr/loc_error_vs_p_Psi_pi_6_T_16_N_T_8_delta_f_120_N_512_6GHz_d_4_Nr_9.dat};


\end{axis}
\end{tikzpicture}\label{fig:Nr_P}}
	\caption{The localization error of the proposed approach for different number of \acp{Rx} in terms of (a): Subcarrier bandwidth. (b): Number of subcarriers. (c): Total transmit power ($N E_s$). The \ac{RIS} location is $[7,  7]$.}
	\label{fig:par_invest_Nr}
	\end{center}
\end{figure}
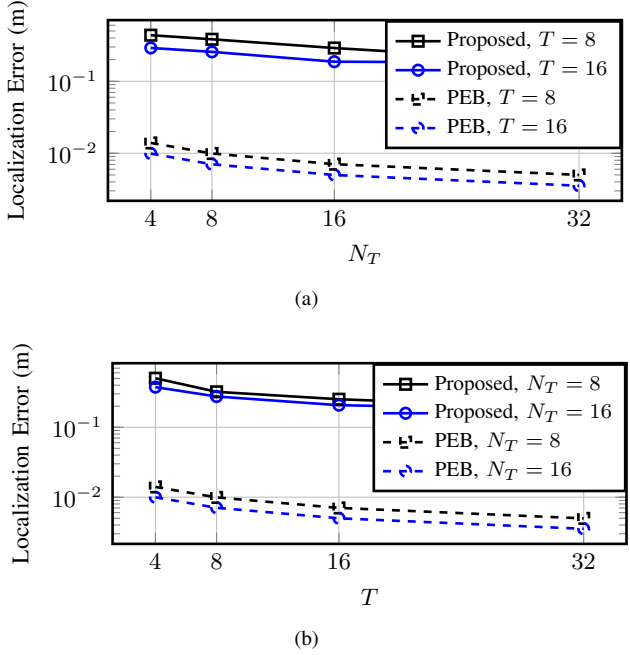
\begin{figure}[t!] 
\begin{center}
	\subfigure[]{\pgfplotsset{every axis/.append style={
		font=\small,
		line width=1pt,
legend style={font=\footnotesize,  at={(1,1)},anchor=north east},legend cell align=left},
} %
\pgfplotsset{compat=1.13}
	\begin{tikzpicture}
\begin{axis}[
xlabel near ticks,
ylabel near ticks,
ymode=log,
grid=major,
xlabel={$N_T$},
ylabel={Localization Error (m) },
	xtick={4,8,16,32},
yticklabel style={/pgf/number format},
width=0.95\linewidth,
height=0.45\linewidth,
	legend columns=1,	
legend entries={
{ Proposed,  $T=8$}, { Proposed,  $T=16$},{ PEB,  $T=8$},{PEB,  $T=16$}
},
ylabel style={font=\small},
xlabel style={font=\small},
]
\definecolor{green-samaneh}{rgb}{0, 0.66, 0.7}
\addplot[black,solid,mark=square] table {Figures/Fig_loc_error_vs_N_T_fix_T/loc_error_vs_N_t_Psi_pi_6_T_8_6GHz_d_4.dat};
\addplot[blue,solid,mark=o] table {Figures/Fig_loc_error_vs_N_T_fix_T/loc_error_vs_N_t_Psi_pi_6_T_16_6GHz_d_4.dat};
\addplot[black,dashed,mark=square] table {Figures/Fig_loc_error_vs_N_T_fix_T/PEB_vs_NT_Psi_pi_6_T_8_delta_f_60_N_512_6GHz_d_4.dat};
\addplot[blue,dashed,mark=o] table {Figures/Fig_loc_error_vs_N_T_fix_T/PEB_vs_NT_Psi_pi_6_T_16_delta_f_60_N_512_6GHz_d_4.dat};

\end{axis}
\end{tikzpicture} \label{fig:vs_NT}}
\subfigure[]{\pgfplotsset{every axis/.append style={
		font=\small,
		line width=1pt,
legend style={font=\footnotesize,  at={(1,1)},anchor=north east},legend cell align=left},
} %
\pgfplotsset{compat=1.13}
	\begin{tikzpicture}
\begin{axis}[
xlabel near ticks,
ylabel near ticks,
ymode=log,
grid=major,
xlabel={$T$},
ylabel={Localization Error (m) },
xtick={4,8,16,32},
yticklabel style={/pgf/number format},
width=0.95\linewidth,
height=0.45\linewidth,
	legend columns=1,	
legend entries={
{ Proposed,  ${N_T}=8$},
{{ Proposed,  ${N_T}=16$}},
{ PEB,  ${N_T}=8$},
{ PEB,  ${N_T}=16$}
},
ylabel style={font=\small},
xlabel style={font=\small},
]
\definecolor{green-samaneh}{rgb}{0, 0.66, 0.7}
\addplot [black,solid,mark=square] table {Figures/Fig_loc_error_vs_T_fix_N_T/loc_error_vs_T_Psi_pi_6_NT_8_6GHz_d_4.dat};
\addplot[blue,solid,mark=o] table {Figures/Fig_loc_error_vs_T_fix_N_T/loc_error_vs_T_Psi_pi_6_NT_16_6GHz_d_4.dat};
\addplot [black,dashed,mark=square] table {Figures/Fig_loc_error_vs_T_fix_N_T/PEB_vs_T_Psi_pi_6_NT_8_delta_f_120_N_512_6GHz_d_4.dat};
\addplot [blue,dashed,mark=o] table {Figures/Fig_loc_error_vs_T_fix_N_T/PEB_vs_T_Psi_pi_6_NT_16_delta_f_120_N_512_6GHz_d_4.dat};



\end{axis}
\end{tikzpicture}\label{fig:vs_T}}
	\caption{{The localization error of the proposed approach in terms of (a): $N_T$  (b): $T$. The \ac{RIS} location is $[7,  7]$.}}
	\label{fig:vs_NT_T}
	\end{center}
\end{figure}
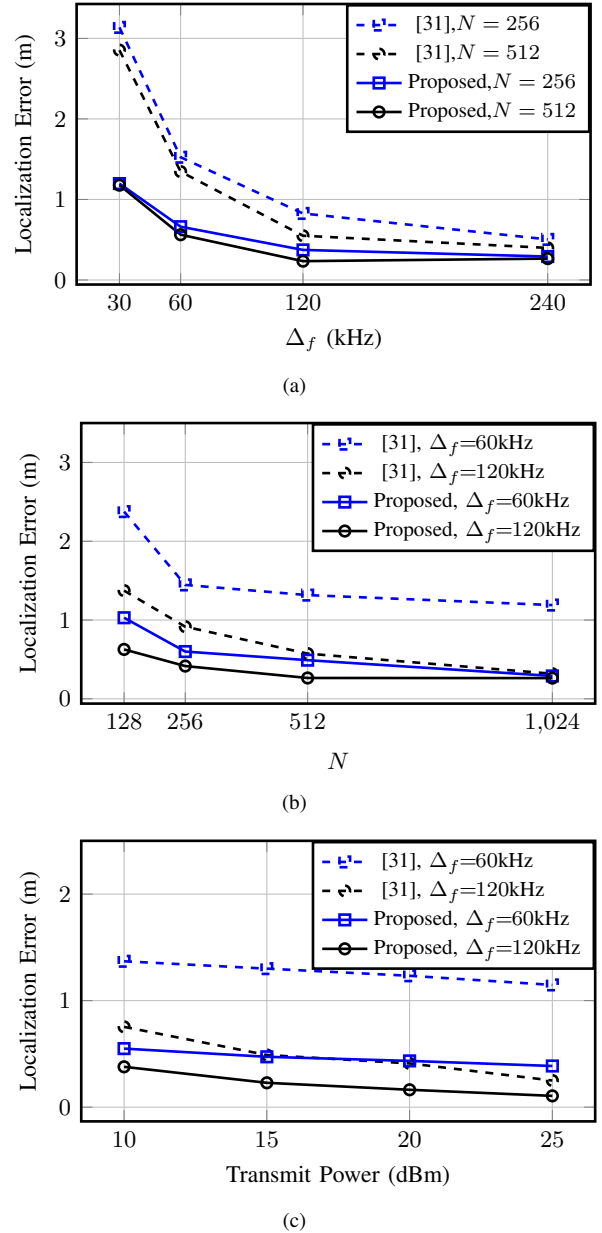
\begin{figure}[t] 
\begin{center}
	\subfigure[]{\pgfplotsset{every axis/.append style={
		font=\small,
		line width=1pt,
legend style={font=\footnotesize,  at={(1,1)},anchor=north east},legend cell align=left},
} %
\pgfplotsset{compat=1.13}
	\begin{tikzpicture}
\begin{axis}[
xlabel near ticks,
ylabel near ticks,
grid=major,
xlabel={$\Delta_f$ (kHz) },
ylabel={Localization Error (m) },
	xtick={30,60,120,240},
yticklabel style={/pgf/number format},
width=0.95\linewidth,
height=0.6\linewidth,
	legend columns=1,	
legend entries={
{\cite{semipassive},$N=256$},
 {\cite{semipassive},$N=512$},
{Proposed,$N=256$}, 
{ Proposed,$N=512$}
},
ylabel style={font=\small},
xlabel style={font=\small},
]
\addplot[blue,dashed,mark=square] table {Figures/Fig_loc_error_vs_delta_f/loc_error_ToA_vs_delta_f_Psi_pi_6_T_16_N_T_8_N_256_6GHz_d_4.dat};
\addplot[black,dashed,mark=o] table {Figures/Fig_loc_error_vs_delta_f/loc_error_ToA_vs_delta_f_Psi_pi_6_T_16_N_T_8_N_512_6GHz_d_4_1.dat};
\addplot[blue,solid,mark=square] table {Figures/Fig_loc_error_vs_delta_f/loc_error_vs_delta_f_Psi_pi_6_T_16_N_T_8_N_256_6GHz_d_4.dat};
\addplot[black,solid,mark=o] table {Figures/Fig_loc_error_vs_delta_f/loc_error_vs_delta_f_Psi_pi_6_T_16_N_T_8_N_512_6GHz_d_4_1.dat};



\end{axis}
\end{tikzpicture} \label{fig:delta_f}}
	\subfigure[]{\pgfplotsset{every axis/.append style={
		font=\small,
		line width=1pt,
			legend style={font=\footnotesize,  at={(1,1)},anchor=north east},legend cell align=left},
} %
\pgfplotsset{compat=1.13}
	\begin{tikzpicture}
\begin{axis}[
xlabel near ticks,
ylabel near ticks,
grid=major,
xlabel={$N$ },
ylabel={Localization Error (m) },
	xtick={128, 256,512,1024,2048},
yticklabel style={/pgf/number format},
width=0.95\linewidth,
height=0.6\linewidth,
	legend columns=1,	
legend entries={
{\cite{semipassive}, $\Delta_f\hspace{-1mm}=\hspace{-1mm}60$kHz},
{\cite{semipassive}, $\Delta_f\hspace{-1mm}=\hspace{-1mm}120$kHz},
 {Proposed, $\Delta_f\hspace{-1mm}=\hspace{-1mm}60$kHz},
{ Proposed, $\Delta_f\hspace{-1mm}=\hspace{-1mm}120$kHz}},
	ymax=3.5,
ylabel style={font=\small},
xlabel style={font=\small},
]
\addplot[blue,dashed,mark=square] table {Figures/Fig_loc_error_vs_N/loc_error_ToA_vs_N_Psi_pi_6_T_16_N_T_8_delta_f_60_6GHz_d_4.dat};
\addplot[black,dashed,mark=o] table {Figures/Fig_loc_error_vs_N/loc_error_ToA_vs_N_Psi_pi_6_T_16_N_T_8_delta_f_120_6GHz_d_4.dat};
\addplot[blue,solid,mark=square] table {Figures/Fig_loc_error_vs_N/loc_error_vs_N_Psi_pi_6_T_16_N_T_8_delta_f_60_6GHz_d_4.dat};
\addplot[black,solid,mark=o] table {Figures/Fig_loc_error_vs_N/loc_error_vs_N_Psi_pi_6_T_16_N_T_8_delta_f_120_6GHz_d_4.dat};


\end{axis}
\end{tikzpicture}\label{fig:N}}
	\subfigure[]{\pgfplotsset{every axis/.append style={
		font=\small,
		line width=1pt,
		legend style={font=\footnotesize,  at={(1,1)},anchor=north east},legend cell align=left},
} %
\pgfplotsset{compat=1.13}
	\begin{tikzpicture}
\begin{axis}[
xlabel near ticks,
ylabel near ticks,
grid=major,
xlabel={Transmit Power (dBm) },
ylabel={Localization Error (m) },
yticklabel style={/pgf/number format},
width=0.95\linewidth,
height=0.6\linewidth,
	legend columns=1,	
legend entries={
{\cite{semipassive}, $\Delta_f\hspace{-1mm}=\hspace{-1mm}60$kHz},
{\cite{semipassive}, $\Delta_f\hspace{-1mm}=\hspace{-1mm}120$kHz},
{ Proposed,  $\Delta_f\hspace{-1mm}=\hspace{-1mm}60$kHz},
{Proposed,  $\Delta_f\hspace{-1mm}=\hspace{-1mm}120$kHz}
},
	ymax=2.5,
ylabel style={font=\small},
xlabel style={font=\small},
]

\addplot[blue,dashed,mark=square] table {Figures/Fig_loc_error_vs_P/loc_error_ToA_vs_p_Psi_pi_6_T_16_N_T_8_delta_f_60_N_512_6GHz_d_4.dat};
\addplot[black,dashed,mark=o] table {Figures/Fig_loc_error_vs_P/loc_error_ToA_vs_p_Psi_pi_6_T_16_N_T_8_delta_f_120_N_512_6GHz_d_4.dat};
\addplot[blue,solid,mark=square] table {Figures/Fig_loc_error_vs_P/loc_error_vs_p_Psi_pi_6_T_16_N_T_8_delta_f_60_N_512_6GHz_d_4.dat};
\addplot[black,solid,mark=o] table {Figures/Fig_loc_error_vs_P/loc_error_vs_p_Psi_pi_6_T_16_N_T_8_delta_f_120_N_512_6GHz_d_4.dat};


\end{axis}
\end{tikzpicture}\label{fig:P}}
	\caption{The localization error in terms of (a): Subcarrier bandwidth. (b): Number of subcarriers. (c): Total transmit power ($N E_s$). The \ac{RIS} location is $[7,  7]$.}
	\label{fig:par_invest}
	\end{center}
\end{figure}
\section{ Proposed Tracking Approach}
\label{sec:tracking}
We propose an \ac{EKF}-based approach for tracking the \acp{RIS}. In this regard, let us define $\mathbf{a}_k$ as the acceleration vector of the $k^{\th}$ \ac{RIS}.  We consider the  fixed acceleration state model for the location and velocity of the \ac{RIS} as follows  
\begin{equation}
{\bf x}_k[n]\hspace{-1.25mm}=\hspace{-1.25mm}\begin{bmatrix}\pk[n]\\
 {\bf v}_k[n]\end{bmatrix}
\hspace{-1.25mm}=\hspace{-1.25mm}\begin{bmatrix}\pk[n-1]\hspace{-0.5mm}+\hspace{-0.5mm}T_s{\bf v}_k[n-1]\hspace{-0.5mm}+\hspace{-0.5mm}\frac{1}{2}T_s^2{\bf a}_k\\
\mathbf{v}_k[n-1]+T_s{\bf a}_k\\ \end{bmatrix}\hspace{-0.5mm}+{\bf u}_{k,1}[n],
\label{eq:state}
\end{equation}
in which ${\bf v}_k[n]$ is the velocity of the \ac{RIS} at the $n^{\th}$ step, $T_s$ is the tracking step time, and ${\bf u}_{k,1}[n]$ is \ac{AWGN}.
 We must note that in the fixed acceleration model, actually the variation of the direction of the RISs is assumed to be fixed. But, the tracking system can adapt itself for the cases that the variation of the direction is not fixed, by updating the value of  acceleration. Thus, the considered model can support paths with different velocities and directions.
Let, define ${\bf A}=\begin{bmatrix}
1&0&T_s&0\\
0&1&0& T_s\\
0&0&1&0\\
0&0&0&1
\end{bmatrix}$ and ${\bf B}=\begin{bmatrix} \frac{T_s^2}{2} & 0 & T_s &0\\
0& \frac{T_s^2}{2}& 0 & T_s
\end{bmatrix}^{T}$, the state model can be rewritten as
\begin{equation}
{\bf x}_k[n]={\bf A}~ {\bf x}_k[n-1]+{\bf B}~{\bf a}_k+ {\bf u}_{k,1}[n].
\end{equation}

Additionally, defining $\xi_{k,n_{\mathrm{r}}}$ as the length of the path between the transmitter and the $n_{\mathrm{r}}^{\th}$ \ac{Rx} including the $k^{\th}$ \ac{RIS}, 
we consider the vector $\bnu_{k}[n]=\left[\hat{\xi}_{k,1}[n],
\dots,
\hat{\xi}_{k,N_{\mathrm{r}}}[n],
\hat{\alpha}_{k,1}[n],
\dots,
\hat{\alpha}_{k,N_{\mathrm{r}}}[n]\right]^T$
as the measurement vector and model it as 
\begin{equation}
\label{eq:measurement}
\bnu_{k}[n]=\underbrace{\scriptsize{\begin{bmatrix}
\left\|\pk[n]-\pt\right\|+\left\|\pk[n]-\prf\right\|\\ 
\vdots
\\ 
\left\|\pk[n]-\pt\right\|+\left\|\pk[n]-\prNr\right\|
\\ \\
\cos \psi_{k,1} \left(\frac{\left|p_{_{k,1}}-p_{_{\mathrm{t},1}}\right|}{\left\|{\mathbf{p}}_{_k}-{\mathbf{p}}_{_{\mathrm{t}}}\right\|}- \frac{\left|p_{_{k,1}}- {p}_{_{{\mathrm{r}}, 1,1}}\right|}{\left\|{\mathbf{p}}_{_{k}}-{\mathbf{p}}_{_{{\mathrm{r}},1}}\right\|}\right)+\\ +\sin \psi_{k,1} \left(\frac{\left|p_{_{k,2}}-p_{_{\mathrm{t},2}}\right|}{\left\|{\mathbf{p}}_{_k}-{\mathbf{p}}_{_{\mathrm{t}}}\right\|}+ \frac{\left|p_{_{k,2}}- {p}_{_{{\mathrm{r}}, 1,2}}\right|}{\left\|{\mathbf{p}}_{_{k}}-{\mathbf{p}}_{_{{\mathrm{r}},1}}\right\|}\right)
\\
\vdots
\\
\cos \psi_{k,N_{\mathrm{r}}} \left(\frac{\left|p_{_{k,1}}-p_{_{\mathrm{t},1}}\right|}{\left\|{\mathbf{p}}_{_k}-{\mathbf{p}}_{_{\mathrm{t}}}\right\|}- \frac{\left|p_{_{k,1}}- {p}_{_{{\mathrm{r}}, N_{\mathrm{r}},1}}\right|}{\left\|{\mathbf{p}}_{_{k}}-{\mathbf{p}}_{_{{\mathrm{r}},N_{\mathrm{r}}}}\right\|}\right)+\\ +\sin \psi_{k,N_{\mathrm{r}}} \left(\frac{\left|p_{_{k,2}}-p_{_{\mathrm{t},2}}\right|}{\left\|{\mathbf{p}}_{_k}-{\mathbf{p}}_{_{\mathrm{t}}}\right\|}+ \frac{\left|p_{_{k,2}}- {p}_{_{{\mathrm{r}}, N_{\mathrm{r}},2}}\right|}{\left\|{\mathbf{p}}_{_{k}}-{\mathbf{p}}_{_{{\mathrm{r}},N_{\mathrm{r}}}}\right\|}\right)
\end{bmatrix}}}_{{\bf h}({\bf x}_k[n])}+{\bf u}_{k,2}[n],
\end{equation}
in which ${\bf u}_{k,2}[n]$ is \ac{AWGN}.

It is noticeable that  the measurement vector is not linear. Hence, we first linearise the equation of the measurement vector,  by defining
\begin{equation}\label{eq:H_n}
{\bf H}_k[n|n-1]=
\dfrac{\partial{\bf h}({\bf x}_k[n])}{\partial{\bf x}_k[n]}\Big|_{{\bf x}_k[n]={{\bf x}}_k[n|n-1]},
\end{equation}
which is obtained in Appendix \ref{app}.

Subsequently, defining ${\bf Q}_k[n]=\mathbb{E}\{{\bf u}_{k,1}[n]{\bf u}_{k,1}[n]^{H}\}$ and ${\bf C}_k[n]=\mathbb{E}\{{\bf u}_{k,2}[n]{\bf u}_{k,2}[n]^{H}\}$ and regarding the \ac{EKF} approach \cite{kay1993fundamentals}, we propose Algorithm \ref{alg2}
to track the location of the \acp{RIS}.
Let, ${\bf M}_k[n|n-1]$, ${\bf K}_k[n]$, and ${\bf M}_k[n|n]$ represent the minimum prediction \ac{MSE}, the Kalman gain, and the minimum \ac{MSE} matrices at the $n^{\th}$ step of the tracking process, respectively.  
 In this algorithm, at the first step, we estimate ${\bf x}_k[0|0]$, the initial location of the $k^{\th}$ \ac{RIS}, using Algorithm \ref{alg1}. Then, at each tracking process, we first predict the location of the \ac{RIS} conditioned on the previous measurements. Next, 
we obtain the minimum prediction \ac{MSE} matrix and use it to calculate the Kalman gain matrix. Then, we correct the location of the RIS employing the Kalman matrix gain and matrix ${\bf H}_k[n]$. Finally, we obtain the minimum \ac{MSE} matrix to be used for the next tracking process.

We can notice that the computational complexity of algorithm \ref{alg2}  considering the number multiplications in estimating the measurement parameters, and the complexity of Kalman filtering, which is of the order of number of the measurement parameters to the power of $2.4$\cite{kay1993fundamentals},   is of the order of $\mathcal{O}(KN_{\mathrm{r}}(NN_T+N_T N_{f,\tau} N_{f,f_{_{\mathrm{d}}}}+N_{f,\tau} N_{f,f_{_{\mathrm{d}}}} \log(N_{f,\tau} N_{f,f_{_{\mathrm{d}}}})+2^{2.4} N_{\mathrm{r}}^{1.4})$.

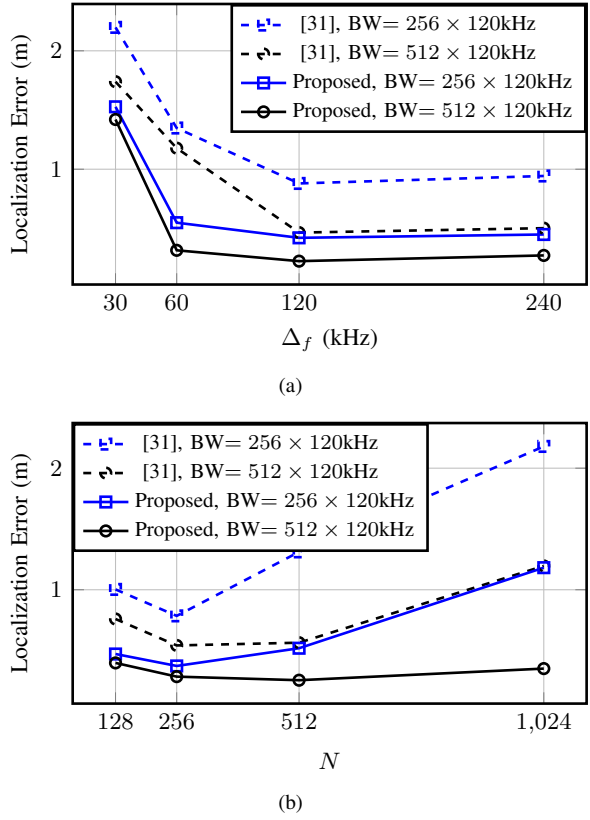
\begin{figure}[t!] 
\begin{center}
	\subfigure[]{\pgfplotsset{every axis/.append style={
		font=\small,
		line width=1pt,
			legend style={font=\footnotesize,  at={(1,1)},anchor=north east},legend cell align=left},
} %
\pgfplotsset{compat=1.13}
	\begin{tikzpicture}
\begin{axis}[
xlabel near ticks,
ylabel near ticks,
grid=major,
xlabel={$\Delta_f$ (kHz) },
ylabel={Localization Error (m) },
	xtick={30, 60,120,240},
yticklabel style={/pgf/number format},
width=0.95\linewidth,
height=0.6\linewidth,
legend entries={
{\cite{semipassive}, BW$=256\times120$kHz},
{\cite{semipassive}, BW$=512\times120$kHz},
 {Proposed,  BW$=256\times120$kHz},
{ Proposed, BW$=512\times120$kHz}},
ylabel style={font=\small},
xlabel style={font=\small},
]
\addplot[blue,dashed,mark=square] table {Figures/Fig_loc_error_vs_delta_f_fix_BW/loc_error_ToA_vs_delta_f_Psi_pi_6_T_16_N_T_8_BW_256_120_6GHz_d_4.dat};
\addplot[black,dashed,mark=o] table {Figures/Fig_loc_error_vs_delta_f_fix_BW/loc_error_ToA_vs_delta_f_Psi_pi_6_T_16_N_T_8_BW_512_120_6GHz_d_4.dat};
\addplot[blue,solid,mark=square] table {Figures/Fig_loc_error_vs_delta_f_fix_BW/loc_error_vs_delta_f_Psi_pi_6_T_16_N_T_8_BW_256_120_6GHz_d_4.dat};
\addplot[black,solid,mark=o] table {Figures/Fig_loc_error_vs_delta_f_fix_BW/loc_error_vs_delta_f_Psi_pi_6_T_16_N_T_8_BW_512_120_6GHz_d_4.dat};


\end{axis}
\end{tikzpicture} \label{fig:delta_f_fix_BW}}
		\subfigure[]{\pgfplotsset{every axis/.append style={
		font=\small,
		line width=1pt,
			legend style={font=\footnotesize,  at={(0,1)},anchor=north west},legend cell align=left},
} %
\pgfplotsset{compat=1.13}
	\begin{tikzpicture}
\begin{axis}[
xlabel near ticks,
ylabel near ticks,
grid=major,
xlabel={$N$ },
ylabel={Localization Error (m) },
	xtick={128, 256,512,1024,2048},
yticklabel style={/pgf/number format},
width=0.95\linewidth,
height=0.6\linewidth,
legend entries={
{\cite{semipassive}, BW$=256\times120$kHz},
{\cite{semipassive}, BW$=512\times120$kHz},
{Proposed,  BW$=256\times120$kHz},
{ Proposed, BW$=512\times120$kHz}},
ylabel style={font=\small},
xlabel style={font=\small},
]
\addplot[blue,dashed,mark=square] table {Figures/Fig_loc_error_vs_N_fix_BW/loc_error_ToA_vs_N_Psi_pi_6_T_16_N_T_8_BW_256_120_6GHz_d_4.dat};
\addplot[black,dashed,mark=o] table {Figures/Fig_loc_error_vs_N_fix_BW/loc_error_ToA_vs_N_Psi_pi_6_T_16_N_T_8_BW_512_120_6GHz_d_4.dat};
\addplot[blue,solid,mark=square] table {Figures/Fig_loc_error_vs_N_fix_BW/loc_error_vs_N_Psi_pi_6_T_16_N_T_8_BW_256_120_6GHz_d_4.dat};
\addplot[black,solid,mark=o] table {Figures/Fig_loc_error_vs_N_fix_BW/loc_error_vs_N_Psi_pi_6_T_16_N_T_8_BW_512_120_6GHz_d_4.dat};


\end{axis}
\end{tikzpicture} \label{fig:N_fix_BW}}
	\caption{{The localization error as the total bandwidth is fixed in terms of (a): Subcarrier bandwidth. (b): Number of subcarriers.  The \ac{RIS} location is $[7,  7]$.}}
	\label{fig:par_invest_fix_BW}
	\end{center}
\end{figure}
\section{Cramer-Rao Lower Bounds}\label{sec:CRLB}
In this section, we obtain the \ac{CRLB} of the location of the $k^{\th}$ \ac{RIS}. Since our approach relies on the \ac{ToA} and geometric parameters,  we also derive the \acp{CRLB} of the \ac{ToA} between the transmitter and the $n_{\mathrm{r}}^{\th}$ receiver through the $k^{\th}$ \ac{RIS}, i.e., $\tau_{k,n_{\mathrm{r}}}$, along with the geometric parameters $\alpha_{k,n_{\mathrm{r}}}$, for $n_{\mathrm{r}}\in \mathcal{N}_{\mathrm{r}}$.  
Considering the original received signal as the observation vector,
we define vector $\mathbf{\bar{y}} \in \mathcal{C}^{N\bar{T} N_{\mathrm{r}} \times 1}$, such that $\left[\mathbf{\bar{y}}\right]_{\gamma_1(n,\bar{t},n_{\mathrm{r}})}=\left[\mathbf{y}_{n_{\mathrm{r}}, \bar{t}}\right]_n$, where $\mathbf{y}_{n_{\mathrm{r}}, \bar{t}}$ is  defined in \eqref{eq:y_t_bar} and $\gamma_1(n,\bar{t},n_{\mathrm{r}})={n+N(\bar{t}-1)+N\bar{T}(n_{\mathrm{r}}-1)}$. 
\vspace{-0.5cm}
\subsection{\ac{CRLB} of $\pk$}
To calculate this, we define 
 \begin{align*} \Beta=&[p_{1,1}, p_{1,2}, g_{1,1}, f_{_{\mathrm{d},1,1}},  \dots, g_{k,N_{\mathrm{r}}}, f_{_{\mathrm{d},1,N_{\mathrm{r}}}}, \dots,
 \\ & p_{K,1}, p_{K,2}, g_{K,1}, f_{_{\mathrm{d},K,1}},  \dots, g_{K,N_{\mathrm{r}}}, f_{_{\mathrm{d},K,N_{\mathrm{r}}}}]^{T}. 
 \end{align*} 
  The vector $\mathbf{\bar{y}}$ can be represented as  $\mathbf{\bar{y}}=\mathbf{f}+\mathbf{z}$ where  
  \begin{align}
\notag & \left[\mathbf{f}\right]_{\gamma_1(n,\bar{t},n_{\mathrm{r}})}=\sum_{k=1}^{K} \sqrt{E}_s \  g_{k,n_{\mathrm{r}}} e^{j 2 \pi (n-1)\Delta f \tau_{k,n_{\mathrm{r}}}} \\ & \times e^{j 2\pi  f_{_{\mathrm{d},k,n_{\mathrm{r}}}} T_{\mathrm{d}} (\bar{t}-1) }  \sum\limits_{m=1}^{M}[\bPhi_{k}(\bar{t})]_{m.m} e^{j\frac{2\pi}{\lambda}(m-1)d\alpha_{k,n_{\mathrm{r}}}}.
  \end{align}
Therefore, the $(i,j)^{\th}$ component of $\mathbf{I}_{\Beta}$, the information matrix of $\Beta$, is equal to \cite{kay1993fundamentals} $$\left[\mathbf{I}_{\Beta}\right]_{i,j}=2 \mathcal{R}\left\{\left[\frac{\partial \mathbf{f}}{\partial \eta_{i}}\right]^{H}\mathbf{Q}_{\mathbf{z}}^{-1}\left[\frac{\partial  \mathbf{f}}{\partial \eta_{j}}\right]\right\},$$
in which $\mathbf{Q}_{\mathbf{z}}$ is the covariance matrix of vector $\mathbf{z}$.
Now, define $\gamma_2(k,n_{\mathrm{r}})=(k-1)n_{\mathrm{r}}$ and
\begin{align*} 
\!\mu(n,k,n_{\mathrm{r}},m) \!\!=\! e^{j 2 \pi( (n-1)\Delta f \tau_{k,n_{\mathrm{r}}}\hspace{-0.25mm}+\hspace{-0.25mm} f_{_{\mathrm{d},k,n_{\mathrm{r}}}} \hspace{-1mm}T_{\mathrm{d}} \hspace{-0.25mm}(\bar{t}-1)+\frac{1}{\lambda}(m-1)d\alpha_{k,n_{\mathrm{r}}})}\hspace{-0.25mm}.
\end{align*}
For $k \in \mathcal{K}$,  $n_{\mathrm{r}}\in \mathcal{N}_{\mathrm{r}}$, and $i^{\prime}\in \{1,2\}$
the element $\gamma_1(n,\bar{t},n_r)$ of 
$\frac{\partial \mathbf{f}}{\partial \eta_{i}}$ for  $i=2\gamma_2(k,N_{\mathrm{r}}+1)+1$ and
$i=2\gamma_2(k,N_{\mathrm{r}}+1)+2$ is as \eqref{eq:partial_f_eta_1_2}, in which
 the partial derivatives of $\alpha_{k,n_{\mathrm{r}}}$ and $\tau_{k,n_{\mathrm{r}}}$, are given as obtained in appendix \ref{app}.  
\begin{figure*}
\begin{align}
\label{eq:partial_f_eta_1_2}
&\left[\frac{\partial \mathbf{f}}{\partial \eta_{2\gamma_2(k,N_{\mathrm{r}}+1)+i^{\prime}}}\right]_{\gamma_1(n,\bar{t},n_{\mathrm{r}})} \!\!\!\!\!\!\!\!\!\!\!\!\!\!\!\!\!\!\!\!= \sqrt{E_s} g_{k,n_{\mathrm{r}}}
\sum\limits_{m=1}^{M}\hspace{-0.5mm}\left[\bPhi_k(\bar{t})\right]_{m,m} \mu(n,k,n_{\mathrm{r}},m) 
\left(j 2 \pi (n-1)\Delta f \frac{\partial \tau_{k,n_{\mathrm{r}}}}{\partial p_{k,i^{\prime}}}+j\frac{2\pi}{\lambda}(m-1)d \frac{\partial \alpha_{k,n_{\mathrm{r}}}}{\partial p_{k,i^{\prime}}}\right),  \ i^{\prime}=1,2. 
\\
\hline \notag
\end{align}
\end{figure*}
Also, for
$i=2\gamma_2(k,N_{\mathrm{r}}+1)+2n_{\mathrm{r}}+1$ the elements of the vector $\frac{\partial \mathbf{f}}{\partial \eta_{i}}$ are equal to  
\begin{align}
\label{eq:partial_f_eta_nr1}
& \!\left[\frac{\partial \mathbf{f}}{\partial \eta_{2\gamma_2(k,N_{\mathrm{r}}+1)+2n_{\mathrm{r}}+1}}\right]_{\!\gamma_1(n,\bar{t},n_{\mathrm{r}})} \!\!\!\!\!\!\!\!\!\!\!\!\!\!\!\!\!\!\!=\hspace{-1mm} \sqrt{E_s}   
\sum\limits_{m=1}^{M} \!\! \left[\bPhi_k(\bar{t})\right]_{m,m} \!\mu(n,k,n_{\mathrm{r}},m),
\end{align}
and for $i=2\gamma_2(k,N_{\mathrm{r}}+1)+2n_{\mathrm{r}}+2$ the elements of this vector are as
\begin{align}
\label{eq:partial_f_eta_nr2}
&\left[\frac{\partial \mathbf{f}}{\partial \eta_{2\gamma_2(k,N_{\mathrm{r}}+1)+2n_{\mathrm{r}}+2}}\right]_{\gamma_1(n,\bar{t},n_{\mathrm{r}})}=  \\ \notag & \sqrt{E_s} g_{k,n_{\mathrm{r}}}  (j2\pi T_{\mathrm{d}} (\bar{t}-1))  
\sum\limits_{m=1}^{M}  \left[\bPhi_k(\bar{t})\right]_{m,m} \mu(n,k,n_{\mathrm{r}},m).
\end{align}
The \ac{CRLB} of $\pk$ is equal  to ${ \rm tr}\left\{\mathbf{J}_{\Beta,2\gamma_2(k,N_{\mathrm{r}}+1)+1:2\gamma_2(k,N_{\mathrm{r}}+1)+2}\right\}$, in which $\mathbf{J}_{\Beta}=\mathbf{I}_{\Beta}^{-1}$ and we have
\begin{align*}&\mathbf{J}_{\Beta, 2\gamma_2(k,N_{\mathrm{r}}+1)+1:2\gamma_2(k,N_{\mathrm{r}}+1)+2}\\ &\hspace{-0.5mm}=\hspace{-1mm}\scriptsize{\begin{bmatrix}
J_{\Beta,2\gamma_2(k,N_{\mathrm{r}}+1)+1,2\gamma_2(k,N_{\mathrm{r}}+1)+1} & J_{\Beta,2\gamma_2(k,N_{\mathrm{r}}+1)+1,2\gamma_2(k,N_{\mathrm{r}}+1)+2} \\ J_{\Beta,2\gamma_2(k,N_{\mathrm{r}}+1)+2,2\gamma_2(k,N_{\mathrm{r}}+1)+1} & J_{\Beta,2\gamma_2(k,N_{\mathrm{r}}+1)+2,2\gamma_2(k,N_{\mathrm{r}}+1)+2}
\end{bmatrix}}.
\end{align*}
 
\subsection{\ac{CRLB} of $\tau_{k,n_{\mathrm{r}}}$ and  $\alpha_{k,n_{\mathrm{r}}}$}
\begin{figure}[t!]
\begin{center}
\subfigure[]{\label{fig:y}\pgfplotsset{every axis/.append style={
		font=\small,
		line width=1pt,
		legend style={font=\footnotesize,  at={(1,1)},anchor=north east},legend cell align=left},
} %
\pgfplotsset{compat=1.13}
	\begin{tikzpicture}
\begin{axis}[
xlabel near ticks,
ylabel near ticks,
grid=major,
xlabel={$y$(m) },
ylabel={Localization Error (m) },
width=0.99\linewidth,
yticklabel style={/pgf/number format},
width=0.95\linewidth,
height=0.6\linewidth,
	legend columns=1,	
legend entries={
{\cite{semipassive}, $x=9$},
{\cite{semipassive}, $x=7$},
{ Proposed, $x=9$},
{Proposed, $x=7$},
},
ylabel style={font=\small},
xlabel style={font=\small},
]
\addplot[black,dashed,mark=o] table {Figures/Fig_loc_error_vs_y/loc_error_ToA_vs_y_Psi_pi_6_T_16_N_T_8_delta_f_120_N_512_x_9_6GHz_d_4.dat};
\addplot[blue,dashed,mark=square] table {Figures/Fig_loc_error_vs_y/loc_error_ToA_vs_y_Psi_pi_6_T_16_N_T_8_delta_f_120_N_512_x_7_6GHz_d_4.dat};
\addplot[black,solid,mark=o] table {Figures/Fig_loc_error_vs_y/loc_error_vs_y_Psi_pi_6_T_16_N_T_8_delta_f_120_N_512_x_9_6GHz_d_4.dat};
\addplot[blue,solid,mark=square] table {Figures/Fig_loc_error_vs_y/loc_error_vs_y_Psi_pi_6_T_16_N_T_8_delta_f_120_N_512_x_7_6GHz_d_4.dat};


\end{axis}
\end{tikzpicture}}
\subfigure[]{\label{fig:x}\pgfplotsset{every axis/.append style={
		font=\small,
		line width=1pt,
		legend style={font=\footnotesize,  at={(1,1)},anchor=north east},legend cell align=left},
} %
\pgfplotsset{compat=1.13}
	\begin{tikzpicture}
\begin{axis}[
xlabel near ticks,
ylabel near ticks,
grid=major,
xlabel={$x$(m) },
ylabel={Localization Error (m) },
yticklabel style={/pgf/number format},
width=0.95\linewidth,
height=0.6\linewidth,
	legend columns=1,	
legend entries={
{\cite{semipassive}, $y=7$},{\cite{semipassive}, $y=4$},
{Proposed,  $y=7$},{Proposed,  $y=4$}},
ylabel style={font=\small},
xlabel style={font=\small},
]
\addplot[black,dashed,mark=o] table {Figures/Fig_loc_error_vs_x/loc_error_ToA_vs_x_Psi_pi_6_T_16_N_T_8_delta_f_120_N_512_y_7_6GHz_d_4.dat};
\addplot[blue,dashed,mark=square] table {Figures/Fig_loc_error_vs_x/loc_error_ToA_vs_x_Psi_pi_6_T_16_N_T_8_delta_f_120_N_512_y_4_6GHz_d_4.dat};
\addplot[black,solid,mark=o] table {Figures/Fig_loc_error_vs_x/loc_error_vs_x_Psi_pi_6_T_16_N_T_8_delta_f_120_N_512_y_7_6GHz_d_4.dat};
\addplot[blue,solid,mark=square] table {Figures/Fig_loc_error_vs_x/loc_error_vs_x_Psi_pi_6_T_16_N_T_8_delta_f_120_N_512_y_4_6GHz_d_4.dat};


\end{axis}
\end{tikzpicture}}
\caption{Localization error in terms of coordinates of the \ac{RIS} (a): $y$. (b): $x$.}
\label{fig:axis}
\end{center}
\end{figure}
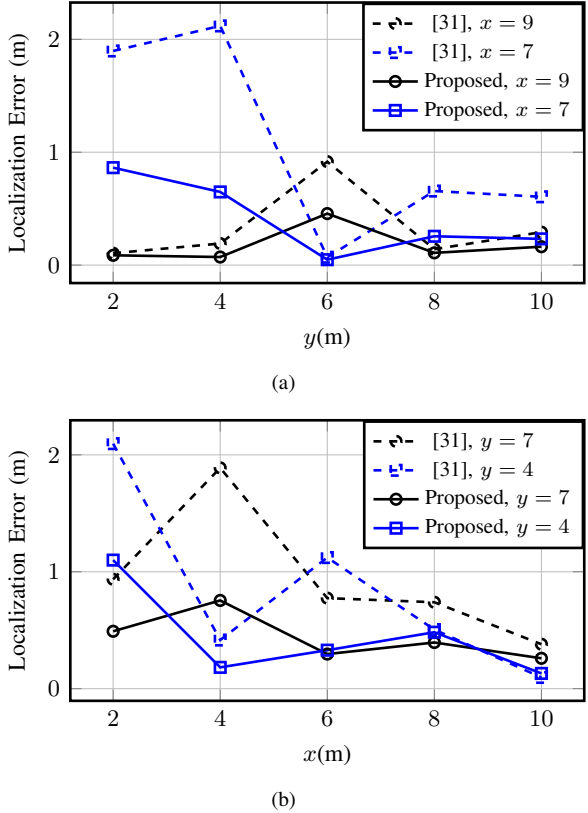
\begin{figure}[t] 
\begin{center}
\includegraphics[width=1\columnwidth]{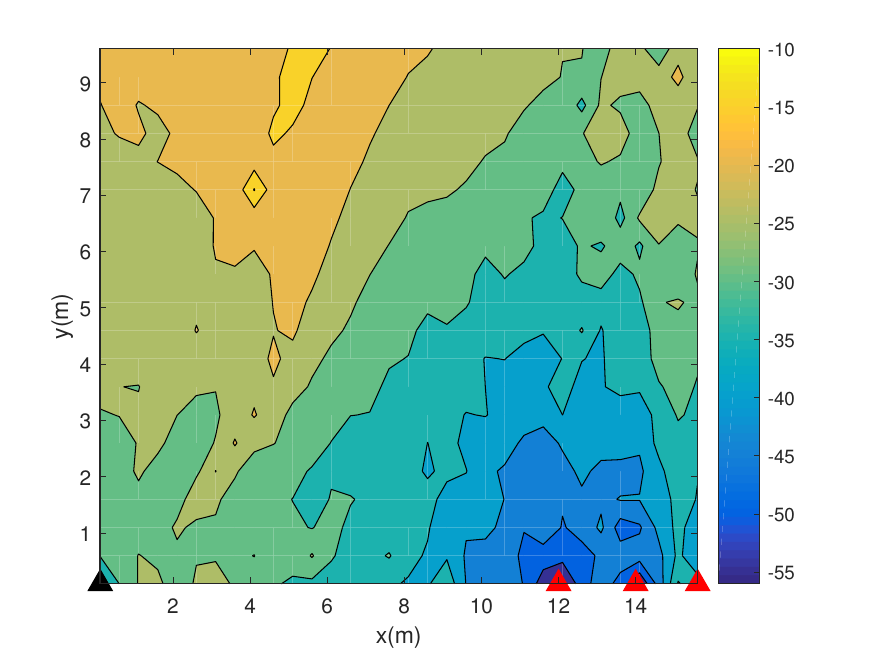}
\end{center}
\caption{ \ac{PEB} (dB) of \ac{RIS} localization using the \ac{Tx} and \acp{Rx}.}
\label{fig:PEB}
\end{figure}
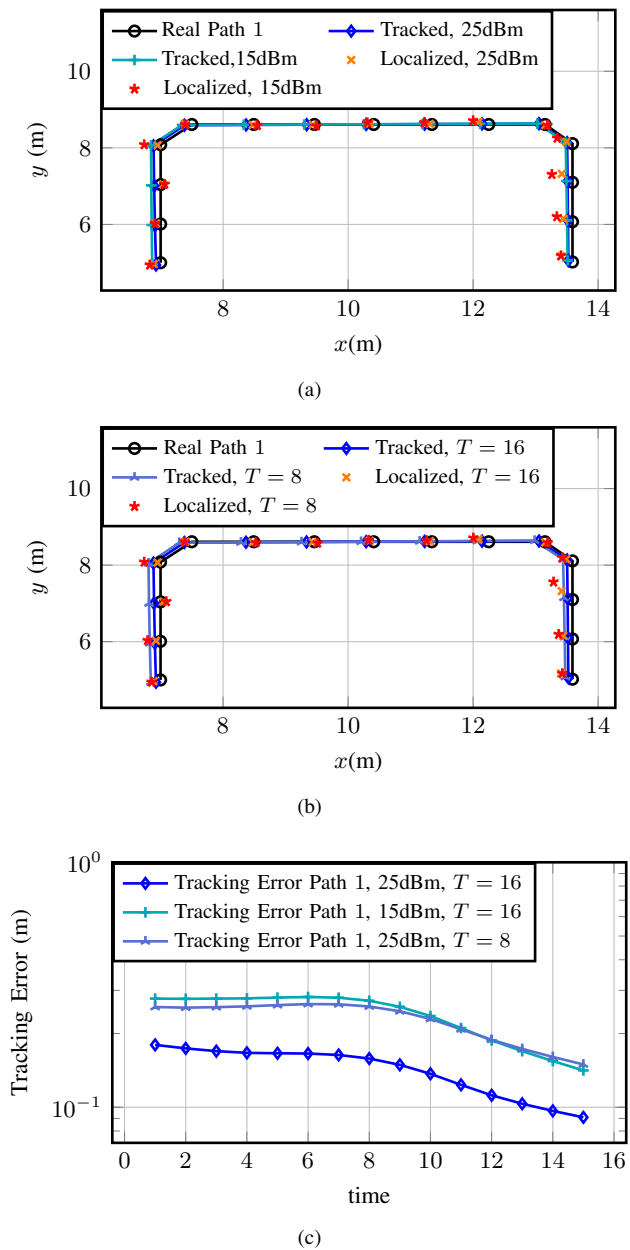
\begin{figure}[t]
\begin{center}
\subfigure[]{\pgfplotsset{every axis/.append style={
		font=\small,
		line width=1pt,
		legend style={font=\footnotesize,  at={(0,1)},anchor=north west},legend cell align=left},
} %
\pgfplotsset{compat=1.13}
	\begin{tikzpicture}
\begin{axis}[
xlabel near ticks,
ylabel near ticks,
grid=major,
xlabel={$x$(m) },
ylabel={$y$ (m) },
width=0.5\linewidth,
yticklabel style={/pgf/number format},
width=0.95\linewidth,
height=0.6\linewidth,
	legend columns=2,	
legend entries={
\hspace{-0.25mm}Real Path $1$,
{Tracked, 25dBm},
{\hspace{-0.25mm}Tracked,15dBm}, { Localized, 25dBm}, { Localized, 15dBm}
},
ymax=11.6,
ylabel style={font=\small},
xlabel style={font=\small},
]
\definecolor{green-samaneh}{rgb}{0, 0.66, 0.7}
\definecolor{midnightgreen}{rgb}{0, 0.29, 0.33}
\definecolor{coolblack}{rgb}{0.0, 0.18, 0.39}
\addplot[black,solid,mark=o] table {Figures/Fig_tracking_paths/tracking_real_path_exact_path1.dat};
\addplot[blue,solid,mark=diamond] table {Figures/Fig_tracking_paths/tracking_tracked_path_exact_path1_N_T_8_T_16_d_4_25dBm.dat};
\addplot[green-samaneh,solid,mark=+] table {Figures/Fig_tracking_paths/tracking_tracked_path_exact_path1_N_T_8_T_16_d_4_15dBm_NT.dat};
\addplot[orange,only marks,mark=x] table {Figures/Fig_tracking_paths/tracked_path_exact_path1_with_localization_15dBm.dat};
\addplot[red,only marks,mark=star] table {Figures/Fig_tracking_paths/tracked_path_exact_path1_with_localization_25dBm.dat};

\end{axis}
\end{tikzpicture}\label{fig:tracking_path_1_power}}
\subfigure[]{\pgfplotsset{every axis/.append style={
		font=\small,
		line width=1pt,
		legend style={font=\footnotesize,  at={(0,1)},anchor=north west},legend cell align=left},
} %
\pgfplotsset{compat=1.13}
	\begin{tikzpicture}
\begin{axis}[
xlabel near ticks,
ylabel near ticks,
grid=major,
xlabel={$x$(m) },
ylabel={$y$ (m) },
width=0.5\linewidth,
yticklabel style={/pgf/number format},
width=0.95\linewidth,
height=0.6\linewidth,
	legend columns=2,	
legend entries={
Real Path $1$,
{Tracked, $T=16$},{ Tracked, $T=8$ }, {Localized, $T=16$},{ Localized, $T=8$}
},
ymax=11.6,
ylabel style={font=\small},
xlabel style={font=\small},
]
\definecolor{green-samaneh}{rgb}{0, 0.66, 0.7}
\definecolor{midnightgreen}{rgb}{0, 0.63, 0.23}
\definecolor{coolblack}{rgb}{0.0, 0.18, 0.39}
\definecolor{bluebell}{rgb}{0.34,0.44, 0.82}
\addplot[black,solid,mark=o] table {Figures/Fig_tracking_paths/tracking_real_path_exact_path1.dat};
\addplot[blue,solid,mark=diamond] table {Figures/Fig_tracking_paths/tracking_tracked_path_exact_path1_N_T_8_T_16_d_4_25dBm.dat};
\addplot[bluebell,solid,mark=Mercedes star] table {Figures/Fig_tracking_paths/tracking_tracked_path_exact_path1_N_T_8_T_8_d_4_25dBm.dat};
\addplot[orange,only marks,mark=x] table {Figures/Fig_tracking_paths/tracked_path_exact_path1_with_localization_15dBm.dat};
\addplot[red,only marks,mark=star] table {Figures/Fig_tracking_paths/tracked_path_exact_path1_with_localization_T_8.dat};

\end{axis}
\end{tikzpicture}\label{fig:tracking_path_1_T}}
\subfigure[]{\pgfplotsset{every axis/.append style={
		font=\small,
		line width=1pt,
		legend style={font=\footnotesize,  at={(0,1)},anchor=north west},legend cell align=left},
} %
\pgfplotsset{compat=1.13}
	\begin{tikzpicture}
\begin{axis}[
xlabel near ticks,
ylabel near ticks,
ymode=log,
grid=major,
xlabel={time},
ylabel={Tracking Error (m) },
yticklabel style={/pgf/number format},
width=0.95\linewidth,
height=0.6\linewidth,
	legend columns=1,	
legend entries={
{Tracking Error Path $1$, 25dBm, $T=16$},{Tracking Error Path $1$, 15dBm,     $T=16$},
{Tracking Error Path $1$, 25dBm, $T=8$}},
ymax=1,
ylabel style={font=\small},
xlabel style={font=\small},
]
\definecolor{green-samaneh}{rgb}{0, 0.66, 0.7}
\definecolor{bluebell}{rgb}{0.34,0.44, 0.82}
\definecolor{zeit}{rgb}{0, 0.63, 0.23}
\definecolor{midnightgreen}{rgb}{0, 0.29, 0.33}
\definecolor{coolblack}{rgb}{0.0, 0.18, 0.39}
\addplot[blue,solid,mark=diamond] table {Figures/Fig_tracking_error/tracking_error_path1_N_T_8_T_16_d_4_25dBm.dat};
\addplot[green-samaneh,solid,mark=+] table {Figures/Fig_tracking_error/tracking_error_path1_N_T_8_T_16_d_4_15dBm_NT.dat};
\addplot[bluebell,solid,mark=Mercedes star] table {Figures/Fig_tracking_error/tracking_error_path1_N_T_8_T_8_d_4_25dBm.dat};
\end{axis}
\end{tikzpicture}\label{fig:error_tracking_path_1}}
\end{center}
\caption{Investigation of tracking accuracy over path 1 (a): For different values of the transmit power. (b):  For different values of $T$ (c): Tracking error over the path.  $\|\mathbf{v}[0]\|=10$m$/$s, $\|\mathbf{a}\|=2$m$/{\mathrm{ s}}^{2}$.}
\label{fig:tracking1}
\end{figure}
\begin{figure}[t!]
\begin{center}
{\pgfplotsset{every axis/.append style={
		font=\small,
		line width=1pt,
		legend style={font=\footnotesize,  at={(1,1)},anchor=north east},legend cell align=left},
} %
\pgfplotsset{compat=1.13}
	\begin{tikzpicture}
\begin{axis}[
xlabel near ticks,
ylabel near ticks,
grid=major,
xlabel={$e (m)$},
ylabel={Average CDF of error},
yticklabel style={/pgf/number format},
width=0.95\linewidth,
height=0.6\linewidth,
	legend columns=1,	
legend entries={$\sigma_0=0$,
$\sigma_0=0.05$,
$\sigma_0=0.2$},
xmin= 0, 
	ymin=0, 
ylabel style={font=\small},
xlabel style={font=\small},
]
\definecolor{green-samaneh}{rgb}{0, 0.66, 0.7}
\definecolor{midnightgreen}{rgb}{0, 0.29, 0.33}
\definecolor{coolblack}{rgb}{0.0, 0.18, 0.39}
\addplot[black,solid,mark=square] table {Figures/Fig_initial_error/CDF_initial_error_path1_sigma2_0.dat};
\addplot[blue,solid,mark=o] table {Figures/Fig_initial_error/CDF_initial_error_path1_sigma2_001.dat};
\addplot[green-samaneh,solid,mark=x] table {Figures/Fig_initial_error/CDF_initial_error_path1_sigma2_02.dat};
\end{axis}
\end{tikzpicture}}
\end{center}
\caption{ Average of the CDFs of error over path 1 for different variances of initial localization error. $\|\mathbf{v}[0]\|=10$m$/$s, $\|\mathbf{a}\|=2$m$/{\rm s}^{2}$. The values of the average error over the path for $\sigma_0=0$, $\sigma_0=0.05$ and $\sigma_0=0.2$ are equal to $8.48 \times 10^{-4}$, $4.38 \times 10^{-2}$ and $0.18$, respectively.  }
\label{fig:initial_CDF}
\end{figure}
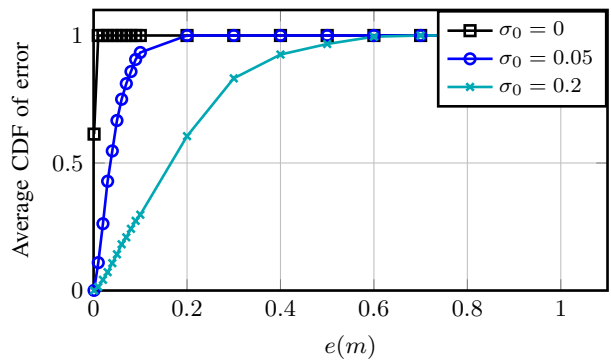
{
To calculate this, let us define 
\begin{align*}
\bbeta =&[\tau_{1,1}, \alpha_{1,1}, g_{1,1}, f_{_{\mathrm{d},1,1}}, \dots \tau_{1,N_{\mathrm{r}}}, \alpha_{1,N_{\mathrm{r}}}, g_{1,N_{\mathrm{r}}},  f_{_{\mathrm{d},1,N_{\mathrm{r}}}}, \dots \\ &\tau_{K,1},  \alpha_{K,1},  g_{K,1},  f_{_{\mathrm{d},K,1}}, \dots, \tau_{K,N_{\mathrm{r}}},  \alpha_{K,N_{\mathrm{r}}}, g_{K,N_{\mathrm{r}}}, f_{_{\mathrm{d},K,N_{\mathrm{r}}}} ]. 
\end{align*}
The $(i,j)^{\th}$ component of $\mathbf{I}_{\bbeta}$, the information matrix of vector  $\bbeta$, is equal to $\left[\mathbf{I}_{\bbeta}\right]_{i,j}=\mathcal{R}\left\{\left[\frac{\partial \mathbf{f}}{\partial \beta_{i}}\right]^{H}\mathbf{Q}_{\mathbf{z}}\left[\frac{\partial  \mathbf{f}}{\partial \beta_{j}}\right]\right\}$.
For $k \in \mathcal{K}$,   and $n_{\mathrm{r}}\in \mathcal{N}_{\mathrm{r}}$,
 the elements of the derivative of vector $\mathbf{f}$ with respect to $\beta_i$   for  $i=4\gamma_2(k,N_{\mathrm{r}})+4n_{\mathrm{r}}-3$ are as  
\begin{align}
\label{eq:par_beta_tau}
& \!\!\!\! \left[\frac{\partial \mathbf{f}}{\partial \beta_{4\gamma_2(k,N_{\mathrm{r}})+4n_{\mathrm{r}}-3}}\right]_{\gamma_1(n,\bar{t},n_{\mathrm{r}})}= \\ \notag &\sqrt{E_s} g_{k,n_{\mathrm{r}}}  (j2 \pi (n-1)\Delta f)   
\sum\limits_{m=1}^{M} \left[\bPhi_k(\bar{t})\right]_{m,m} \mu(n,k,n_{\mathrm{r}},m) \notag.
\end{align}
Moreover, the elements of $\frac{\partial \mathbf{f}}{\partial \beta_{i}}$ for
$i=4\gamma_2(k,N_{\mathrm{r}})+4n_{\mathrm{r}}-2$ 
are equal to
\begin{align}
\label{eq:par_beta_alpha}
& \left[\frac{\partial \mathbf{f}}{\partial \beta_{4\gamma_2(k,N_{\mathrm{r}})+4n_{\mathrm{r}}-2}}\right]_{\gamma_1(n,\bar{t},n_{\mathrm{r}})}= \\ & \notag 
j\frac{2\pi}{\lambda} d \sqrt{E_s} g_{k,n_{\mathrm{r}}} \sum\limits_{m=1}^{M}(m-1) \left[\bPhi_k(\bar{t})\right]_{m,m} \mu(n,k,n_{\mathrm{r}},m),
\end{align}
and for
$i=4\gamma_2(k,N_{\mathrm{r}})+4n_{\mathrm{r}}-1$ and $i=4\gamma_2(k,N_{\mathrm{r}})+4n_{\mathrm{r}}$ are given by (\ref{eq:par_beta_g}) and (\ref{eq:par_beta_f_d}), respectively.
\begin{align}
\label{eq:par_beta_g}
& \left[\frac{\partial \mathbf{f}}{\partial \beta_{4\gamma_2(k,N_{\mathrm{r}})+4n_{\mathrm{r}}-1}}\right]_{\gamma_1(n,\bar{t},n_{\mathrm{r}})} \!\!\!\!\!\!\!\!\!\!\!\!\!\!\!\!\!\!\!\!= \sqrt{E_s}   
\sum\limits_{m=1}^{M} \!\! \left[\bPhi_k(\bar{t})\right]_{m,m}  \mu(n,k,n_{\mathrm{r}},m),
\end{align}
\begin{align}
\label{eq:par_beta_f_d}
& \left[\frac{\partial \mathbf{f}}{\partial \beta_{4\gamma_2(k,N_{\mathrm{r}})+4n_{\mathrm{r}}}}\right]_{\gamma_1(n,\bar{t},n_{\mathrm{r}})}= \\ & \sqrt{E_s} g_{k,n_{\mathrm{r}}} (j2\pi T_{\mathrm{d}} (\bar{t}-1))  
\sum\limits_{m=1}^{M}  \left[\bPhi_k(\bar{t})\right]_{m,m} \mu(n,k,n_{\mathrm{r}},m). \notag
\end{align}
The \acp{CRLB} of $\tau_{k,n_{\mathrm{r}}}$ and $\alpha_{k,n_{\mathrm{r}}}$ are respectively equal to $[\mathbf{J}_{\bbeta}]_{4\gamma_2(k,N_{\mathrm{r}})+4n_{\mathrm{r}}-3, 4\gamma_2(k,N_{\mathrm{r}})+4n_{\mathrm{r}}-3}$ and $[\mathbf{J}_{\bbeta}]_{4\gamma_2(k,N_{\mathrm{r}})+4n_{\mathrm{r}}-2, 4\gamma_2(k,N_{\mathrm{r}})+4n_{\mathrm{r}}-2}$ in which $\mathbf{J}_{\bbeta}=\mathbf{I}_{\bbeta}^{-1}$.
\section{Simulation Results}
\label{sec:simulation}
In this Section, we provide the simulation results to demonstrate the performance of the proposed localization and tracking approaches. We consider a \ac{Tx} located at $\pt=[0, 0]$ and three \acp{Rx} located at $\mathbf{p}_{_{ \mathrm{r}, 1}}=[ 12, 0 ]$, $\mathbf{p}_{_{ \mathrm{r}, 2}}=[14,  0]$ and $\mathbf{p}_{_{ \mathrm{r}, 3}}=[16,  0]$, respectively. In each simulation two RISs are considered such that the location of one of them is as explained in the caption of the figures, while the location of the other RIS is random and the results are averaged on this location.
The other simulation settings are as shown in Table \ref{tab:sim_parameters}. 
In the following, we present the simulations of the localization and tracking approaches, each in an individual  subsection. 
For the localization performance we use the approach of \cite{semipassive} as a benchmark. To have a fair comparison, our proposed phase shift design approach is used for both the localization approaches and the number of the transmissions are the same. 

\begin{table}[!t]
\scriptsize
\caption{Simulation parameters }
\label{tab:sim_parameters}
\vspace{-2mm}
\centering
\begin{tabular}{|l|c|}
\hline
{ \textbf{Parameter}}								& { \textbf{Value}} 					\\ \hline\hline
Carrier Frequency	($f_{c}$)				& $ 6 $GHz				\\ \hline
\ac{RIS} element spacing ($d$) &  $\lambda/4 $                                                               \\ \hline
Total Transmit power ($N E_s$)							& $30$dBm				\\ \hline
  Number of  subcarriers   ($N$) & $512$ \\ \hline
    Subcarrier bandwidth ($\Delta_f$) &  $120$ kHz    \\ \hline
     $T$ & $32$ \\ \hline
     $N_T$ & $2$ \\ \hline
  \ac{FFT} dimensions  ($N_{f,\tau}$) & $8192$ \\ \hline
    \ac{FFT} dimensions  ($N_{f, f_{_{\mathrm{d}}}}$) & $1024$ \\ \hline
  The angle of orientation of the RIS ($\psi_{k,n_{\mathrm{r}}}$) &   $\pi/6$ \\ \hline
$N_0$					& $ -174\  \text{dBm/Hz} $			\\ \hline
\end{tabular}
\end{table}

\subsection{Localization Performance}
In this Section, we investigate the performance of the proposed localization approach.
\subsubsection{Effect of Number of \acp{Rx}}
We  explore the effect of the number of \acp{Rx} on the performance of the proposed localization approach as depicted in Fig. \ref{fig:par_invest_Nr}. In this figure, we compare the localization error of the systems with three, five, and nine \acp{Rx} in terms of $\Delta_f$, $N$ and total transmit power. As expected, we observe that  increasing the number of \acp{Rx} enhances the localization accuracy. Obviously, this improvement is resulted by leveraging a larger set of measurements made available by larger number of Rxs. 
\subsubsection{Effect of Parameters $N_T$ and $T$}
In Fig. \ref{fig:vs_NT_T}, we verify the impact of the parameters $N_T$ and $T$ on the localization error of the proposed approach and \ac{PEB}. In Fig.  \ref{fig:vs_NT}, we observe that increasing the number of transmissions by increasing $N_T$ reduces the error of the localization approach as well as the value of \ac{PEB}. It is noticeable that increasing $N_T$ actually improves the accuracy of the estimation of $\alpha_k$ (see \eqref{eq:LS_alpha}).
In Fig. \ref{fig:vs_T}, we notice that increasing $T$ also improves the accuracy of the localization approach and the \ac{PEB}, which is due to the effect of increasing $T$ on the variance of the noise of the observation vector.  However, we must note that enhancing the performance of the localization approach by increasing values of $N_T$ and $T$ will be at the cost of increasing the delay and power consumption for localization. Also, it is notable  that the discrepancy between the \ac{PEB} and the localization error stems from the fact that the \ac{CRLB} does not consider the available resolution and cares more about the accuracy relying the assumption that the problem is resolvable. However, in the proposed localization approach we utilize the \ac{FFT} method as a practical estimator for the \ac{ToA} and the Doppler shift, which is constrained by the resolution of the \ac{FFT} operation, which depends on the bandwidth.
\subsubsection{Effect of Parameters $\Delta_f$, $N$ and Total Transmit Power}
In Fig. \ref{fig:par_invest}, we investigate the effect of subcarrier bandwidth ($\Delta_f$), number of subcarriers ($N$) and total transmit power on the performance of the proposed localization approach and compare the results with  \cite{semipassive}. In Fig. \ref{fig:delta_f}, we illustrate the localization error in terms of $\Delta_f$ for different values of $N$. 
We observe that increasing the subcarrier bandwidth,  enhances the accuracy of both the localization approaches. Indeed, by  increasing the value of $\Delta_f$ the accuracy of \acp{ToA} estimation is improved, leading to better localization results. Also, we can  note that the proposed approach has significantly enhanced the localization accuracy in comparison with \cite{semipassive}. Specifically, in this case, our approach achieves the same localization accuracy, but at a half number of subcarriers.
 In Fig. \ref{fig:N}, we plot the localization error in terms of $N$. We observe that increasing the number of subcarriers leads to a reduction in the localization error for both the proposed approaches.
  We can note that in the case of the proposed localization approach, increasing the number of subcarriers actually enhances the  estimation of $\alpha_{k,n_{\mathrm{r}}}$ as well as the estimation of $\tau_{k,n_{\mathrm{r}}}$ (see \eqref{eq:s_hat} and Algorithm 1).
In Fig.\ref{fig:P}, we depict the localization error in terms of total transmit power for different subcarrier bandwidths. We observe that as expected, increasing the transmit power reduces the error of both the localization approaches due to the improvement of the \ac{SNR} of the received signals. Moreover, for a  localization accuracy of $0.1$m, our proposed localization approach reduces the transmit power roughly by 20$\%$.

In Fig. \ref{fig:par_invest_fix_BW}, we investigate the effect of $\Delta_f$ and $N$ as the total bandwidth is fixed. We depict the localization error in terms of $\Delta_f$ and $N$,  for different bandwidth values BW$=256\times 120$kHz and BW$=512\times 120$kHz. 
We observe that for both the localization approaches, there is a optimal value for $\Delta_f$ and $N$.

\subsubsection{Effect of the Position of the \ac{RIS}}
To further investigate the performance of the proposed approach,  in Fig. \ref{fig:axis}, we examine its performance in terms of the $x$-coordinate and $y$-coordinate of the \ac{RIS} location. In Fig. \ref{fig:y}, it is evident that as the distance between the \ac{RIS} and the line between the \ac{Tx} and \ac{Rx} increases, the superiority of the proposed approach over \cite{semipassive} becomes more pronounced. 
Also, Fig. \ref{fig:x} shows the improvement of the accuracy by the proposed approach for different values of $x$. The behaviour of the curves in terms of $x$ and $y$ can be attributed to the \ac{PEB} of the localization in various points of the region. In this regard, we plot the heat map of the \ac{PEB} in Fig. \ref{fig:PEB}. We observe that as the distance between the \ac{RIS} and the line between the \ac{Tx} and \ac{Rx} increases, the \ac{PEB} tends to increase. Moreover, the minimum \ac{PEB} occurs near the \acp{Rx}, while near the \ac{Tx} the error is  higher.
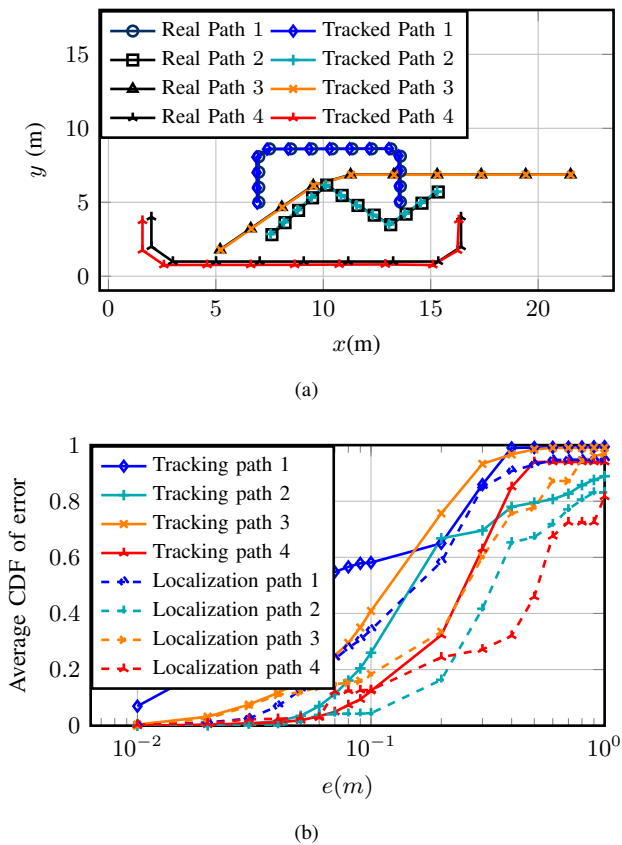
\begin{figure}[t!]
\begin{center}
\subfigure[]{\pgfplotsset{every axis/.append style={
		font=\small,
		line width=1pt,
		legend style={font=\footnotesize,  at={(0,1)},anchor=north west},legend cell align=left},
} %
\pgfplotsset{compat=1.13}
	\begin{tikzpicture}
\begin{axis}[
xlabel near ticks,
ylabel near ticks,
grid=major,
xlabel={$x$(m) },
ylabel={$y$ (m) },
yticklabel style={/pgf/number format},
width=0.95\linewidth,
height=0.6\linewidth,
	legend columns=2,	
legend entries={
Real Path $1$,
Tracked Path $1$, Real Path $2$,
Tracked Path $2$, Real Path $3$,
Tracked Path $3$, Real Path $4$,
Tracked Path $4$},
ymax=18,
ylabel style={font=\small},
xlabel style={font=\small},
]
\definecolor{green-samaneh}{rgb}{0, 0.66, 0.7}
\definecolor{midnightgreen}{rgb}{0, 0.29, 0.33}
\definecolor{coolblack}{rgb}{0.0, 0.18, 0.39}
\addplot[coolblack,solid,mark=o] table {Figures/Fig_tracking_paths/tracking_real_path_exact_path1.dat};
\addplot[blue,solid,mark=diamond] table {Figures/Fig_tracking_paths/tracking_tracked_path_exact_path1_N_T_8_T_16_d_4_25dBm.dat};
\addplot[black,solid,mark=square] table {Figures/Fig_tracking_paths/tracking_real_path_exact_path2.dat};
\addplot[green-samaneh,solid,mark=+] table {Figures/Fig_tracking_paths/tracking_tracked_path_exact_path2_N_T_8_T_16_d_4_25dBm.dat};
\addplot[black,solid,mark=triangle] table {Figures/Fig_tracking_paths/tracking_real_path_exact_path3_v0_20.dat};
\addplot[orange,solid,mark=x] table {Figures/Fig_tracking_paths/tracking_tracked_path_exact_path3_N_T_8_T_16_d_4_v0_20_25dBm.dat};
\addplot[black,solid,mark=Mercedes star] table {Figures/Fig_tracking_paths/tracking_real_path_exact_path4_v0_20.dat};
\addplot[red,solid,mark=Mercedes star] table {Figures/Fig_tracking_paths/tracking_tracked_path_exact_path4_N_T_8_T_16_d_4_v0_20_25dBm.dat};



\end{axis}
\end{tikzpicture}\label{fig:tracking_path123}}
\end{center}
\begin{center}
\subfigure[]{\pgfplotsset{every axis/.append style={
		font=\small,
		line width=1pt,
		legend style={font=\footnotesize,  at={(0,1)},anchor=north west},legend cell align=left},
} %
\pgfplotsset{compat=1.13}
	\begin{tikzpicture}
\begin{axis}[
xlabel near ticks,
ylabel near ticks,
xmode=log,
grid=major,
xlabel={$e (m)$},
ylabel={Average CDF of error},
yticklabel style={/pgf/number format},
width=0.95\linewidth,
height=0.6\linewidth,
	legend columns=1,	
legend entries={ Tracking path 1, Tracking path 2, Tracking path 3, Tracking path 4, Localization path 1, Localization path 2,  Localization path 3, Localization path 4},
xmin= 0, 
xmax=1,
	ymin=0, 
ymax=1,
ylabel style={font=\small},
xlabel style={font=\small},
]
\definecolor{green-samaneh}{rgb}{0, 0.66, 0.7}
\definecolor{midnightgreen}{rgb}{0, 0.29, 0.33}
\definecolor{coolblack}{rgb}{0.0, 0.18, 0.39}
\addplot[blue,solid,mark=diamond] table {Figures/Fig_tracking_error/CDF_avg_path1_N_T_8_T_16_d_4_25dBm.dat};
\addplot[green-samaneh,solid,mark=+] table {Figures/Fig_tracking_error/CDF_avg_path2_N_T_8_T_16_d_4_25dBm.dat};
\addplot[orange,solid,mark=x] table {Figures/Fig_tracking_error/CDF_avg_path3_N_T_8_T_16_d_4_v0_20_25dBm.dat};
\addplot[red,solid,mark=Mercedes star] table {Figures/Fig_tracking_error/CDF_avg_path4_N_T_8_T_16_d_4_v0_20_25dBm.dat};
\addplot[blue,dashed,mark=diamond] table {Figures/Fig_tracking_error/CDF_loclization_path1_T_16_25dBm.dat};
\addplot[green-samaneh,dashed,mark=+] table {Figures/Fig_tracking_error/CDF_loclization_path2_25dBm.dat};
\addplot[orange,dashed,mark=x] table {Figures/Fig_tracking_error/CDF_loclization_path3_T_16_25dBm_v0_20.dat};
\addplot[red,dashed,mark=Mercedes star] table {Figures/Fig_tracking_error/CDF_loclization_path4_T_16_25dBm_v0_20.dat};
\end{axis}
\end{tikzpicture}\label{fig:track_loc_error}}
\end{center}
\caption{Investigation of tracking accuracy, (a): Tracked paths. (b): Average CDF of the error of the proposed tracking and localization approach over the paths.  For paths 1 and 2 $\|\mathbf{v}[0]\|=10$m$/$s, for paths 3 and 4 $\|\mathbf{v}[0]\|=20$m$/$s, and for all paths $\|\mathbf{a}\|=2$m$/{\mathrm{s}}^{2}$. The values of the average error of the tracking approach over the paths are equal to $0.14$, $0.26$, $0.16$ and $0.43$, respectively. Also, the values of the average error of the localization approach over the paths are equal to $0.21$, $0.32$, $0.42$ and $0.66$, respectively.}
\label{fig:tracking2}
\end{figure}
\subsection{Tracking Performance}
Now, we investigate the performance of the proposed tracking approach. In this regard, we examine the tracking of path 1 in Fig. \ref{fig:tracking1}. In Fig. \ref{fig:tracking_path_1_power} and Fig. \ref{fig:tracking_path_1_T}, we plot the tracked path using different values of $T$ and transmit power, and also using the proposed localization approach.  We observe that both increasing $T$ and the transmit power could improve the accuracy of the tracking approach. Also, comparing the tracking and localization approach we can observe that in most points the accuracy of the tracking approach is higher. Moreover, the benefit of the tracking approach compared to the localization is that in tracking the error at each point is less compared to the previous points.  In Fig. \ref{fig:error_tracking_path_1}, we illustrate the error of the tracking approach over path 1. We observe that as expected, for different values of $T$ and transmit power the tracking approach reduces the error of the localization over the time. 

In Fig. \ref{fig:initial_CDF}, we investigate the effect of the initial localization error on the error of tracking over the path.  In this Fig., we illustrate the average CDF of the error over the path for different values of standard deviation of the initial localization error and also obtain the average error over the path. We observe that as expected the accuracy of the tracking error is affected by the variance of the initial error, but for each variance the tracking error reduces the average error compared to the error of the initial point.

To further investigate the performance of the tracking approach, in Fig. \ref{fig:tracking2}, we examine the tracking approach for two different initial velocities, and also compare the accuracy of the tracking over each path with accuracy of the localization approach. The initial velocity of the RIS on paths 1 and 2 is $10 {\rm m/s}$, while its initial velocity on paths 3 and 4 is $20 {\rm m/s}$. 
In Fig. \ref{fig:tracking_path123}, we observe that in all the cases the tracking approach compensates  the initial error along the path. 
 In Fig. \ref{fig:track_loc_error}, we compare the average CDFs of the error of the tracking  and the localization approaches over the paths. We observe that on average over each path the tracking approach could increase the probability of having less error, for different values of $e$, compared  to the localization approach. 
\section{Conclusion}
\label{sec:conclusion}
In this paper, we investigated the problem of tracking \acp{RIS} employing a single antenna transmitter and multiple single antenna receivers.  In this regard, we proposed a phase shift design approach to eliminate the effects of scatters  and extract the signal passed through each \ac{RIS}. Then, we proposed a localization approach to estimate the initial location of the \acp{RIS}. The simulation results demonstrate that this approach outperforms the previously proposed \ac{ToA}-based approach. Finally, based on the \ac{EKF} approach we proposed a tracking algorithm. The simulation results confirm the accuracy of the proposed tracking approach. 
\begin{figure*}[h]
\begin{align}  
&\notag [{\bf H}_k[n|n-1]]_{ N_{\mathrm{r}}+n_{\mathrm{r}},1}=\notag \cos \psi_{k,n_{\mathrm{r}}}\Big( \frac{\left(p_{k,1}[n|n-1]-\ptOne\right)}{\left|p_{k,1}[n|n-1]-\ptOne\right| \left\|\pk[n|n-1]-\pt\right\|} -\frac{\left(p_{k,1}[n|n-1]-\ptOne\right)\left|p_{k,1}[n|n-1]-\ptOne\right|}{ \left\|\pk[n|n-1]-\pt\right\|^3}\vspace{1mm} \\ & -\frac{\left(p_{k,1}[n|n-1]-\prOne\right)}{\left|p_{k,1}[n|n-1]-\prOne\right| \left\|\pk[n|n-1]-\pr\right\|} +\frac{\left(p_{k,1}[n|n-1]-\prOne\right)\left|p_{k,1}[n|n-1]-\prOne\right|}{ \left\|\pk[n|n-1]-\pr\right\|^3}\Big)+ \notag
\\ 
&\sin \psi_{k,n_{\mathrm{r}}} \Big(-\frac{\left|p_{k,2}[n|n-1]-\ptTwo\right|\left(p_{k,1}[n|n-1]-\ptOne\right)}{\left\|\pk[n|n-1]-\pt\right\|^3} -\frac{\left|p_{k,2}[n|n-1]-\prTwo\right|\left(p_{k,1}[n|n-1]-\prOne\right)}{\left\|\pk[n|n-1]-\pr\right\|^3}\Big).
\label{eq:grad_h_1_N_r_n_r} \tag{35}
\\
&  \notag [{\bf H}_k[n|n-1]]_{N_{\mathrm{r}}+n_{\mathrm{r}},2}= \cos \psi_{k,n_{\mathrm{r}}}\Big(-\frac{\left|p_{k,1}[n|n-1]-\ptOne\right|\left(p_{k,2}[n|n-1]-\ptTwo\right)}{\left\|\pk[n|n-1]-\pt\right\|^3} + \\ & \notag \frac{\left|p_{k,1}[n|n-1]-\prOne\right|\left(p_{k,2}[n|n-1]-\prTwo\right)}{\left\|\pk[n|n-1]-\pr\right\|^3}\Big)+ \sin \psi_{k,n_{\mathrm{r}}}\Big( \frac{\left(p_{k,2}[n|n-1]-\ptTwo\right)}{\left|p_{k,2}[n|n-1]-\ptTwo\right| \left\|\pk[n|n-1]-\pt\right\|} \\ \notag &-\frac{\left(p_{k,2}[n|n-1]-\ptTwo\right)\left|p_{k,2}[n|n-1]-\ptTwo\right|}{ \left\|\pk[n|n-1]-\pt\right\|^3}\vspace{1mm}  +\frac{\left(p_{k,2}[n|n-1]-\prTwo\right)}{\left|p_{k,2}[n|n-1]-\prTwo\right| \left\|\pk[n|n-1]-\pr\right\|} \\ & -\frac{\left(p_{k,2}[n|n-1]-\prTwo\right)\left|p_{k,2}[n|n-1]-\prTwo\right|}{ \left\|\pk[n|n-1]-\pr\right\|^3}\Big)
\label{eq:grad_h_2_N_r_n_r} \tag{36}.
\\ \notag
\\
\hline 
\notag
\end{align}
\end{figure*}
\appendices
\section{}
\label{app}
In this appendix,  we derive the partial derivatives of $\tau_{k,n_{\mathrm{r}}}$ and $\alpha_{k,n_{\mathrm{r}}}$, and using them  obtain the elements of matrix ${\bf H}_k[n|n-1]$. The partial derivative of $\tau_{k,n_{\mathrm{r}}}$ with respect to $p_{k,i}$ for $i=1,2$ is equal to
\begin{align}
\frac{\partial \tau_{k,n_{\mathrm{r}}}}{\partial p_{k,i}}=\frac{p_{k,i}-p_{{\rm t},i}}{c \left\|\pk-\pt\right\|}+\frac{p_{k,i}-p_{{\rm r},n_{\mathrm{r}},i}}{c \left\|\pk-\pr\right\|}, \ i=1,2,
\end{align}
and the partial derivatives of $\alpha_{k,n_{\mathrm{r}}}$ with respect to $p_{k,1}$ and $p_{k,2}$ are respectively given as
\begin{align}
\notag &\frac{\partial \alpha_{k,n_{\mathrm{r}}}}{\partial p_{k,1}}=\cos \psi_{k,n_{\mathrm{r}}} \left(\frac{\left(p_{k,1}-\ptOne\right)}{\left|p_{k,1}-\ptOne\right| \left\|\pk-\pt\right\|}\right. \\ &-\frac{\left(p_{k,1}-\ptOne\right)\left|p_{k,1}-\ptOne\right|}{ \left\|\pk-\pt\right\|^3}-\frac{\left(p_{k,1}-\prOne\right)}{\left|p_{k,1}-\prOne\right| \left\|\pk-\pr\right\|}+\notag \\ & \left. \frac{\left(p_{k,1}-\prOne\right)\left|p_{k,1}-\prOne\right|}{ \left\|\pk-\pr\right\|^3}\right)+\notag \\ & \sin \psi_{k,n_{\mathrm{r}}} \left(\frac{\left(p_{k,2}-\ptTwo\right)}{\left|p_{k,2}-\ptTwo\right| \left\|\pk-\pt\right\|}-\right. \notag \\ & \frac{\left(p_{k,2}-\ptTwo\right)\left|p_{k,2}-\ptTwo\right|}{ \left\|\pk-\pt\right\|^3}+ \frac{\left(p_{k,2}-\prTwo\right)}{\left|p_{k,2}-\prTwo\right| \left\|\pk-\pr\right\|} \notag\\ & \left.  -\frac{\left(p_{k,2}-   \prTwo\right)\left|p_{k,2}-\prTwo\right|}{ \left\|\pk-\pr\right\|^3}\right) ,
\end{align}
and
\begin{align}
&\notag \frac{\partial \alpha_{k,n_{\mathrm{r}}}}{\partial p_{k,2}}=-\frac{\left|p_{k,2}-\ptTwo\right|\left(p_{k,1}-\ptOne\right)}{\left\|\pk-\pt\right\|^3}\\ &-\frac{\left|p_{k,2}-\prTwo\right|\left(p_{k,1}-\prOne\right)}{\left\|\pk-\pr\right\|^3}.
\end{align}

Thus, considering \eqref{eq:H_n},  the elements $(n_{\mathrm{r}},i)$ of ${\bf H}_k[n|n-1]$ for $n_{\mathrm{r}}\in \mathcal{N}_{\mathrm{r}}$ and $i=1,\dots,4$ would be as follows
\begin{align}
& [{\bf H}_k[n|n-1]]_{n_{\mathrm{r}},1}=\frac{p_{k,1}[n|n-1]-\ptOne}{ \left\|\pk[n|n-1]-\pt\right\|}\\ \notag&+\frac{p_{k,1}[n|n-1]-\prOne}{\left\|\pk[n|n-1]-\pr\right\|},
 \end{align}
\begin{align}
&[{\bf H}_k[n|n-1]]_{n_{\mathrm{r}},2}=\frac{p_{k,2}[n|n-1]-\ptTwo}{ \left\|\pk[n|n-1]-\pt\right\|} \\ \notag &+\frac{p_{k,2}[n|n-1]-\prTwo}{ \left\|\pk[n|n-1]-\pr\right\|}, 
 \end{align}
\begin{align}
&[{\bf H}_k[n|n-1]]_{ n_{\mathrm{r}},3}=T_s\frac{p_{k,1}[n|n-1]-\ptOne}{ \left\|\pk[n|n-1]-\pt\right\|}\\ \notag&+T_s \frac{p_{k,1}[n|n-1]-\prOne}{ \left\|\pk[n|n-1]-\pr\right\|},
  \end{align}
\begin{align}
&[{\bf H}_k[n|n-1]]_{n_{\mathrm{r}},4}=T_s \frac{p_{k,2}[n|n-1]-\ptTwo}{\left\|\pk[n|n-1]-\pt\right\|} \\ \notag &+T_s\frac{p_{k,2}[n|n-1]-\prTwo}{ \left\|\pk[n|n-1]-\pr\right\|}.
\end{align}
Additionally, the elements $( n_{\mathrm{r}}+N_{\mathrm{r}},i)$ of ${\bf H}_k[n|n-1]$ for  $n_{\mathrm{r}} \in \mathcal{N}_{\mathrm{r}}$ and $i=1,\dots,4$ are as \eqref{eq:grad_h_1_N_r_n_r}, \eqref{eq:grad_h_2_N_r_n_r}, \eqref{eq:grad_h_3_N_r_n_r} and \eqref{eq:grad_h_4_N_r_n_r}.
\begin{align}
&[{\bf H}_k[n|n-1]]_{ N_{\mathrm{r}}+n_{\mathrm{r}},3}= T_s [{\bf H}_k[n|n-1]]_{ N_{\mathrm{r}}+n_{\mathrm{r}},1}.\label{eq:grad_h_3_N_r_n_r}\tag{37}
\\ 
\label{eq:grad_h_4_N_r_n_r}\tag{38}
& [{\bf H}_k[n|n-1]]_{N_{\mathrm{r}}+n_{\mathrm{r}},4}=  T_s[{\bf H}_k[n|n-1]]_{N_{\mathrm{r}}+n_{\mathrm{r}},2}.
\end{align}

\bibliographystyle{IEEEtran}
\bibliography{IEEEabrv,Elzanaty_bibliography.bib}
\begin{IEEEbiography}[{\includegraphics[width=1in,height=1.25in,clip,keepaspectratio]{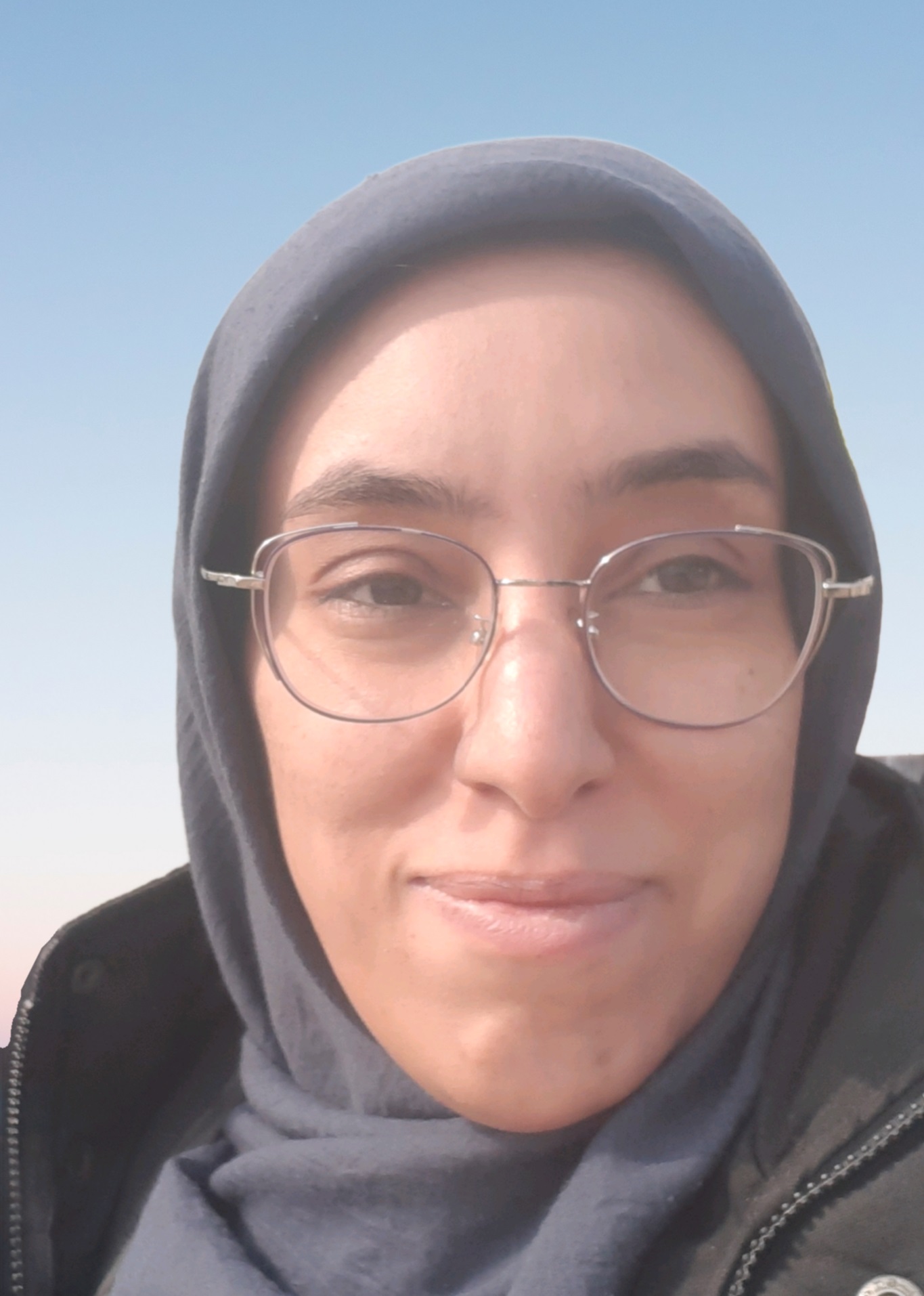}}]{Somayeh Aghashahi}
(S'16) received the B.Sc. degree in pure mathematics from  Shahid Bahonar University of Kerman, Kerman, Iran, in 2013, 
and  received the M.Sc. degree in communication systems engineering from  Yazd Unversity, Yazd, Iran,
 in 2018, where she is currently pursuing the Ph.D degree in communication systems engineering. 
Her  research interests include wireless communication and statistical signal processing
with a particular emphasis on MIMO communications, RIS and reconfigurable antennas assisted
 communication and localization. 
\end{IEEEbiography}
\begin{IEEEbiography}[{\includegraphics[width=1in,height=1.25in,clip,keepaspectratio]{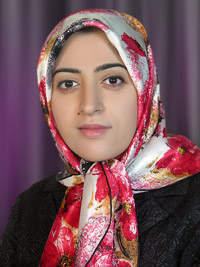}}]
    {Zolfa Zeinalpour-Yazdi} received the B.S., M.S., and Ph.D. degrees in electrical engineering
from the Sharif University of Technology, Tehran, Iran, in 2002, 2004, and 2010, respectively.
She was a Visiting Researcher with ftw. Telecommunication Research Center, Vienna, Austria, in 2006. She joined Yazd University in 2010, where she is currently an Associate Professor with the Department of Electrical Engineering. Her research interest includes the broad area of wireless communications and currently, she is working towards next generation of wireless networks and mathematical treatment of the problems in these areas with a particular
emphasis on stochastic analysis of heterogeneous networks, vehicular communications, RIS-empowered transmissions and also interference management and caching strategy in these networks.
\end{IEEEbiography}
\begin{IEEEbiography} [{\includegraphics[width=1in,height=1.25in,clip,keepaspectratio]{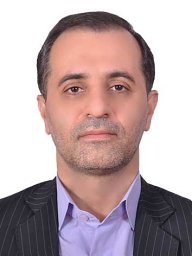}}]
{Aliakbar Tadaion} (S' 05-M' 07-SM' 13) was born in Iran in 1976. He received the B.Sc. degree in electronics, the M.Sc. and Ph.D. degrees in communication systems all from Sharif University of Technology, Tehran, Iran, in 1998, 2000 and 2006, respectively.  
From August 2004 to June 2005, he was with the Department of Electrical and Computer Engineering, Queen's University, Kingston, ON, Canada, as the visiting researcher. He joined the department of electrical Engineering, Yazd University, Yazd, Iran in Sep. 2005, where he is now a Professor.  He is also the founder and director of Cell Lab. 
Now he is a senior member of the Institute of Electrical and Electronics Engineers (IEEE) and also since Jan. 2013, he has been a member of the excom committee of IEEE Iran Section and the newsletter editor of the Iran Section. He has been the technical program committee member of some conferences, such as WOSSPA 2011, IWCIT 2013-2024 and has served as an organizing committee member of some conferences such as ISCC 2012.
\end{IEEEbiography}
\begin{IEEEbiography}[{\includegraphics[width=1in,height=1.25in,clip,keepaspectratio]{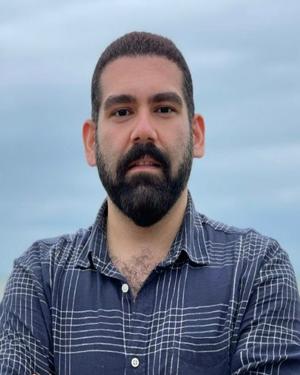}}]{Mahdi Boloursaz Mashhadi }(S’15-M’19-SM’23) is a Lecturer at the 5G/6G Innovation Centre (5G/6GIC) at the Institute for Communication Systems (ICS), University of Surrey (UoS), and a Surrey AI fellow. His research is focused at the intersection of AI/ML with wireless communication, learning and communication co-design, generative AI for telecommunications, and collaborative machine learning. He received B.S., M.S., and Ph.D. degrees in mobile telecommunications from the Sharif University of Technology (SUT), Tehran, Iran. He has more than 40 peer reviewed publications and patents in the areas of wireless communications, machine learning, and signal processing. He is a PI/Co-PI for various government and industry funded projects including the UKTIN/DSIT 12M£ national project TUDOR. He received the Best Paper Award from the IEEE EWDTS conference, and the Exemplary Reviewer Award from the IEEE ComSoc in 2021 and 2022. He served as a panel judge for the International Telecommunication Union (ITU) on the ``AI/ML in 5G" challenge 2021-2022. He is an associate editor for the Springer Nature Wireless Personal Communications Journal.
\end{IEEEbiography}
\begin{IEEEbiography}
[{\includegraphics[width=1in,height=1.25in,clip,keepaspectratio]{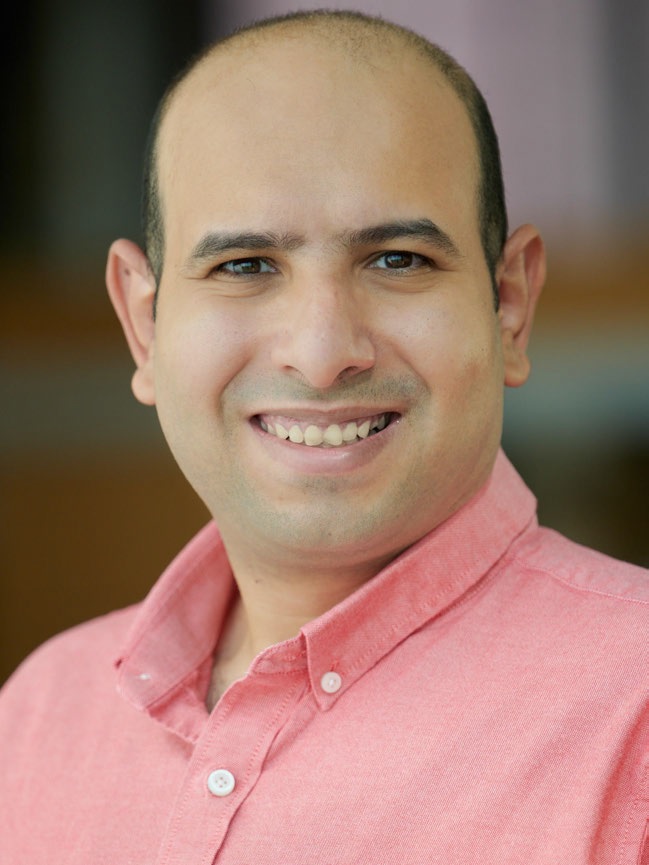}}]
{Ahmed Elzanaty} (S’13-M’18-SM’22) received his PhD degree (cum laude) in electronics, telecommunications, and information technology from the University of Bologna, Italy, in 2018. Currently, he is a Lecturer (Assistant Professor) with the Institute for Communication Systems, University of Surrey, U.K. Before that, he was a Postdoctoral Fellow at the King Abdullah University of Science and Technology (KAUST), Saudi Arabia.  He has participated in several national and European projects, such as TUDOR, CHEDDAR, and EuroCPS. His research interests include the design and performance analysis of wireless communications and localization systems.
\end{IEEEbiography}

\end{document}